\documentclass[aps,prd,twocolumn,superscriptaddress,groupedaddress,amsmath,amssymb,nofootinbib]{revtex4}  
\usepackage{graphicx}
\usepackage{dcolumn}   
\usepackage{bm}        
\usepackage{amssymb}   
\newcommand{\st}{\scriptsize}
\usepackage{bm}
\usepackage[usenames ,dvipsnames]{xcolor}
\usepackage{slashed}
\usepackage{lipsum}
\usepackage{color}
\usepackage{url}
\usepackage{stfloats}
\usepackage{dsfont}        
\usepackage{slashed}
\usepackage{epsfig}
\usepackage{multirow,array}
\usepackage{float}
\usepackage{fancyhdr}
\usepackage{indentfirst}
\usepackage{cancel}
\usepackage{tabularx}
\usepackage{simplewick}
\usepackage{amsmath}
\usepackage{slashed}
\usepackage[normalem]{ulem}
\RequirePackage[colorlinks=true
  ,urlcolor=Green
  ,anchorcolor=blue
  ,citecolor=orange
  ,filecolor=blue
  ,linkcolor=blue
  ,menucolor=blue
  ,pagecolor=blue
  ,linktocpage=true
  ,pdfproducer=medialab
  ,pdfa=true
]{hyperref}

\def\beq{\begin{equation}}
\def\eeq{\end{equation}}
\def\bea{\begin{eqnarray}}
\def\eea{\end{eqnarray}}
\begin{document}


\title{New Perspectives on Axion Misalignment Mechanism}
     
\author{Chia-Feng Chang}
\email{chiafeng.chang@email.ucr.edu}
\author{Yanou Cui} 
\email{yanou.cui@ucr.edu}                                                      
\affiliation{Department of Physics and Astronomy, University of California, Riverside, CA 92521, USA}

\date{\today}

\begin{abstract}
A zero initial velocity of the axion field is assumed in the conventional misalignment mechanism. We propose an alternative scenario where the initial velocity is nonzero, which may arise from an explicit breaking of the PQ symmetry in the early Universe. We demonstrate that, depending on the specifics of the initial velocity and the time order of the PQ symmetry breaking vs. inflation, this new scenario can either enhance or suppress the axion relic abundance relative to the conventional prediction. As a result, new viable parameter regions for axion dark matter may open up.

\end{abstract}
\maketitle

\section{Introduction}
Axions are ultra-light pseudo-scalar particles that are generically predicted in the Peccei-Quinn (PQ) mechanism \cite{Peccei:1977hh,Peccei:1977ur,Wilczek:1977pj}, a compelling solution for the Strong CP problem in particle physics. Recently QCD axions and axion-like particles (ALPs) have attracted substantial interest as a leading dark matter (DM) candidate alternative to WIMPs \cite{Weinberg:1977ma,Abbott:1982af,Dine:1982ah,Preskill:1982cy}.

 \indent Understanding the production mechanism of axions is critical for determining their potential as a viable DM candidate and related phenomenology \cite{Irastorza:2018dyq,Marsh:2015xka}. Despite extensive literature on this subject, our understanding is not yet complete. For instance, for post-inflationary PQ symmetry breaking, axion topological defects (cosmic strings, domain walls) necessarily form and through their subsequent decays may contribute to axion relic abundance ($\Omega_a$) in significant ways \cite{Sikivie:1982qv,Vilenkin:1984ib,Davis:1986xc,Vincent:1996rb,Kawasaki:2014sqa,Vilenkin:1986ku}. However, the prediction of such contributions is still challenging, while a growing effort has recently been made \cite{Klaer:2017qhr,Gorghetto:2018myk,Kawasaki:2018bzv,Martins:2018dqg,Buschmann:2019icd,Hindmarsh:2019csc,Martins:2018,Hook:2018dlk}. Meanwhile, our understanding of possible outcomes of the misalignment production may not be complete either. According to the conventional misalignment mechanism, axion field starts with an initial value, $\theta_i$ ($\theta\equiv a/Nf_a$), away from the true vacuum, then begins to oscillate around the minimum when its mass $m_a\sim3H$ ($H$: Hubble expansion rate), and behaves like cold DM after that. In order to solve the equation of motion for axion evolution, the initial velocity $\dot{\theta}_i$ also needs to be specified, which is implicitly assumed to be zero in the conventional misalignment, and directly affects the $\Omega_a$ prediction. Meanwhile, nonzero $\dot{\theta}_i$ is possible and well-motivated. \textit{Then how would a nonzero initial velocity of the axion field influence the axion relic abundance and phenomenology?}  
 
\indent In this work we propose and systematically investigate an alternative misalignment mechanism with an initial condition $\dot{\theta}_i\neq0$. Based on classified benchmark examples in a UV model independent approach \footnote{We gave an example UV model in Appendix.~B.}, we demonstrate the conditions when axion relic density prediction can significantly differ from the conventional, with potentially dramatic enhancement or suppression depending on specifics with $\dot{\theta}_i$ and whether the PQ symmetry breaks before/during or after inflation. An example model realizing such an initial condition is illustrated in Appendix.~B. Another recent work \cite{Co:2019jts} also considered the possibility of $\dot{\theta}_i\neq0$, focusing on the large $\dot{\theta}>0$ region, demonstrating examples of interesting UV complete models leading to an enhanced $\Omega_a$.

\section{The Origin of a Nonzero Initial Velocity}
Axion originates from the phase of a complex scalar $\Phi$ whose vacuum expectation value $f_a/\sqrt{2}$ leads to the spontaneous breaking of the $U(1)_{\hbox{\st{PQ}}}$ symmetry (or a generic global $U(1)$ for ALPs) \footnote{Without loss of generality we focus on the simplest scenario where domain wall number $N=1$.}:
\begin{align}
\label{Eq:PQscalar}
\Phi \equiv \frac{1}{\sqrt{2}} \left( f_a + \phi \right)e^{i a/f_a}
\end{align}
where $\phi$ and $a$ are the radial and angular (axion) modes, respectively. The conserved Noether charge associated with the PQ symmetry is $R^3 f_a^2 \dot{\theta}$ where $R$ is the cosmic scale factor. $\dot{\theta}_i\neq0$ thus corresponds to the rotation of the $\Phi$ field and an asymmetry of the global PQ charge. Such a charge asymmetry can result from higher dimensional operators that explicitly breaks $U(1)_{\hbox{\st{PQ}}}$ in the early Universe, in analogy to the Affleck-Dine (AD) mechanism for baryogenesis \cite{Affleck:1984fy,Dine:1995uk,Dine:1995kz,Kamionkowski:1992mf}. This effect in fact can be generic for an approximate global symmetry \cite{Kamionkowski:1992mf,Dine:1995uk,Dine:1995kz}. Alternatively $\dot{\theta}_i\neq0$ may originate from axion models with a small dimensionful symmetry-breaking term which introduces a slope in axion potential \cite{Graham:2015cka}. Such PQ-breaking effects should be absent today in order not to undermine the solution to the Strong CP problem, which can be realized by tying its strength to the Hubble rate or a dynamical field that has a larger value in the early Universe. Although the specifics of $\dot{\theta}_i$ is model-dependent, important phenomenological insights can be obtained by studying the axion evolution with benchmark examples that we will demonstrate.\\

\section{Axion Misalignment Mechanism with a Nonzero Initial Velocity}
\begin{figure}[h]
\includegraphics[width=0.46\textwidth]{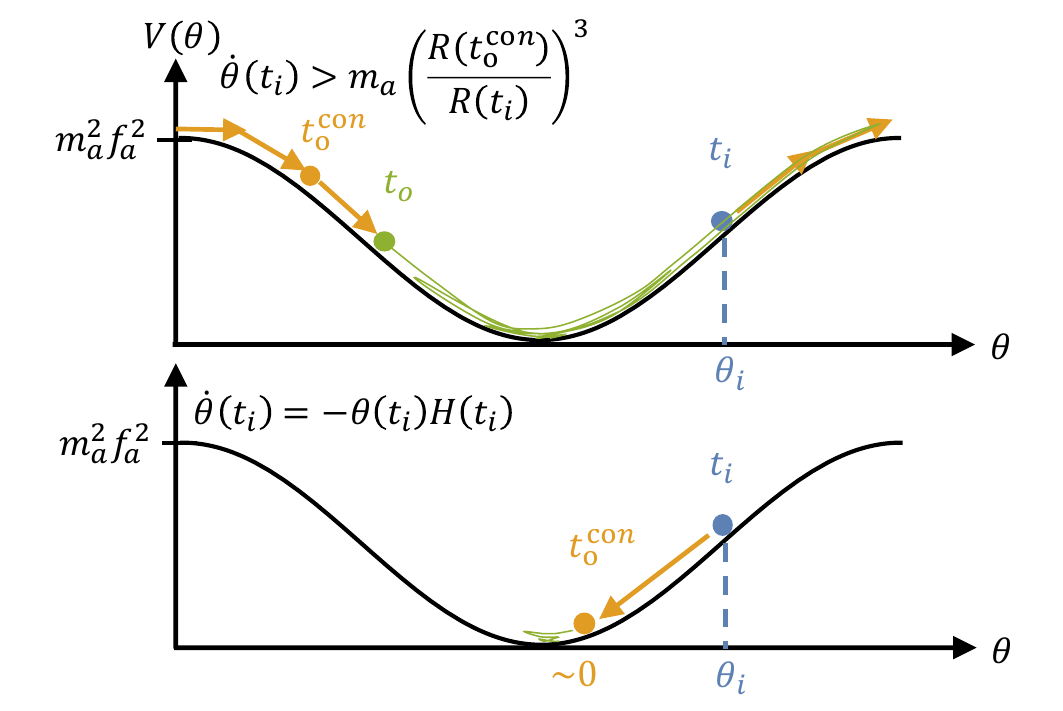} 
\caption{\label{F1} Cartoon illustration of the axion field evolution for the two representative possibilities with $\dot{\theta}_i\neq0$, as explained in the text. The axion starts at $\theta_i$ as blue, then follows the orange arrows until it starts to oscillate at $t_o(t^{con}_o)$. The green trajectory represents the sequence of motion. In the conventional misalignment, the field starts with $\dot{\theta}_i=0$. }
\end{figure}

We first present Fig.~\ref{F1} as a cartoon illustration for two representative possibilities of $\dot{\theta}_i \neq0$ initial condition (IC), with related technical details elaborated later. In conventional misalignment, axion field starts at rest with the rescaled field value $\theta_i$, then roll down the periodic potential well, and start oscillating when $m_a\sim3H$ (at $t_o^{con} \sim 1/m_a$). In both cases we show, the field starts with $\theta_i$ at time $t_i$. The lower panel demonstrates the possibility where the axion field has a negative moderate initial velocity to allow it to roll down further in the potential well so that the field value becomes smaller than $\theta_i$ when oscillation begins. This would lead to a suppression of the axion relic density. The upper panel shows the possibility with a high initial velocity, which may delay the onset of oscillation (at $t_o$) to be later than $t^{con}_o$, thus reduce the entropy $s(t_o)$ then and enhance the relic density of the axion ($\Omega_a\propto Y_a\equiv n_a/s$).\\
We now study the dynamics of axion evolution in details. The equation of motion (e.o.m) of axion field with rescaled $\theta(t) \equiv a(t)/f_a$ (mod $2\pi$) in FRW cosmology is \footnote{We assume that by $t_i$ of our consideration the radial mode has settled into its true vacuum, therefore does not influence axion evolution through coupled e.o.m's. }
\begin{align}
\label{Eq:EoM}
\ddot{\theta} + 3 H \dot{\theta} + m_a^2(T) \theta=0. 
\end{align}
In the conventional misalignment mechanism, $\dot{\theta}(t_i)\equiv\dot{\theta}_i = 0$, and the axion field freezes at a random initial field value $\theta_i$ from PQ breaking to QCD phase transition. For post-inflation PQ breaking, $\Omega_a$ is obtained by averaging over the randomly distributed $\theta_i$ over all causal patches. For QCD axion, we will assume $m_a(T)\propto T^{-4}$ as found by instanton calculation \cite{diCortona:2015ldu, Bonati:2018blm, Petreczky:2016vrs, Burger:2018fvb, Gorghetto:2018ocs}, while a constant $m_a$ may apply for general ALPs. \\

We start by investigating the axion evolution at early times well before oscillation starts. The starting time $t_i$ is generally assumed to be at PQ breaking scale, but can be later times when the axion picks up a nonzero $\dot{\theta}_i$. Allowing $\dot{\theta} (t_i) \neq0$, and dropping the potential term  in Eq.~\ref{Eq:EoM} which is negligible for this early regime, we find the following solution in general cosmology, with background energy density $\rho\propto R^{-n}$:
\begin{align} 
\label{eq: theta_sol}
\theta(t) =&\,\left\{           
\begin{aligned}
 &\,\theta_i + \frac{\dot{\theta}_i}{H_i} \left( \frac{2}{6-n}\right) \left[1-\left( \frac{R(t)}{R(t_i)} \right)^{n/2-3} \right], \; (n\neq 6)\\
&\, \theta_i + \frac{\dot{\theta}_i}{3H_i} \hbox{ln}\left[ \frac{t}{t_i} \right] , \;\;\;\;\;\;\;\;\;\;\;\;\;\;\;\;\;\;\;\;\;\;\;\;\;\;\;\;\;\;\;\;\;\;\;\;\;\, (n= 6)
\end{aligned}
\right.\\
\dot{\theta}(t) =&\, \dot{\theta_i} \left( \frac{R(t_i)}{R(t)} \right)^3.
\end{align}
We will assume standard cosmology (i.e. radiation domination (RD), $n=4$, thus $\theta(t)\propto1/\sqrt{t}$) except when considering inflationary effects. The energy density of the axion evolves as $\rho_a(t)=\frac{1}{2}\dot{\theta}^2(t)f_a^2+\frac{1}{2}m_a(T)^2\theta(t)^2f_a^2$, and relic density can be estimated as: 
\beq
\Omega_a= m_a(T_o)m_a(T=0)\theta_o^2f_a^2\frac{s_0}{{s(t_o)}\rho_c},
\eeq
where $\theta_o\equiv\theta(t_o)$, $t_o$ implicitly depends on $\dot{\theta}_i$, $s_0$ and $\rho_c$ are the current-day entropy and critical density, respectively.\\
We now specify two benchmark types of initial condition of $\dot{\theta}_i$ to find concrete form of solutions given by Eqs.~\ref{eq: theta_sol}.\\
\underline{\textbf{Type-I IC: $\dot{\theta}_i=-\delta H_i$,}} where $\delta$ is a constant parameter independent of ${\theta}_i$. To simplify discussion we choose the convention of $\delta\geq0$ without loss of generality \footnote{The late time evolution of the field with $\delta<0$ is nearly the same as their $\delta>0$ counterpart.}. While the detailed UV physics leading to such initial conditions is not our focus here, they may arise from an AD-like scenario \cite{Dine:2003ax}, which readily gives $|\dot{\theta}(t_i)| \sim H(t_i)$ upon PQ-breaking at $t_i \sim m_{\hbox{\st{pl}}}/f_a^2$ (assuming PQ-breaking during radiation domination around $T\sim f_a$) \footnote{The specifics of initial velocity depends on the radial mode in the UV model. The AD mechanism can generate a initial velocity with a varying ratio to the Hubble rate, see e.g. \cite{Dine:2003ax, Akita:2017ecc,Akita:2018zma}.}. Applying this IC to Eq.~\ref{eq: theta_sol} in RD we find
\begin{align}\label{eq: sol_IC1}
\theta(t)= \theta_i-\delta +\delta \left( \frac{R(t_i)}{R(t)}\right),~~{\rm Mod}[2\pi].
\end{align}
Provided a small/moderate $\dot{\theta}(t_i)$, the oscillation onset $t_o$ in the new scenario is also close to $t_o^{con}$. However, with sufficiently large $\dot{\theta}_i$ the kinetic energy (KE) could be larger than the potential energy V at $t_o^{con}$, and oscillation can only start later when KE $\sim$ V. Such a delayed $t_o$ may enhance $\Omega_a$ since  $\Omega_a\sim \rho_a(t_o)/s(t_o)$. For high $|\dot{\theta_i}|$ we can estimate $t_o$ as the earliest time when KE and $V$ become equal ($ \theta_o\approx2\pi$): 
\beq\label{eq: tosc}
\dot{\theta}_i\left( t_i/t_o \right)^{3/2} \approx \, 2 \pi \, m_a (t_o).
\eeq
Combining Eq.~\ref{eq: tosc} and $m_a(t_o^{con})\approx3H(t_o^{con})$, we can find the critical $\delta_c$: for $\delta>\delta_c$ a notable delay of $t_o$ relative to $t_o^{con}$ would occur: 
\begin{align}
\label{Eq: delta_c}
\delta_c \approx 6\pi  \left({t_o^{con}}/{t_i} \right)^{1/2} \simeq 2 \times 10^{11} \left( \frac{f_a}{10^{11}\,\hbox{GeV}} \right)^{7/6},
\end{align}
where we assumed $t_i$ around the PQ-breaking time $\sim M_{P}/f_a^2$. Note that $\delta_c$ can be much smaller if $t_i$ (when axion picks up a nonzero velocity) is much later than PQ-breaking time.\\

\indent Next we will discuss the evolution patterns in different $\delta$ parameter regions based on Eqs.~\ref{eq: theta_sol}, \ref{eq: sol_IC1}. The two distinct scenarios of PQ breaking after and before/during inflation will be considered in order.\\
\twocolumngrid
\begin{figure}[h] 
\includegraphics[width=0.45\textwidth]{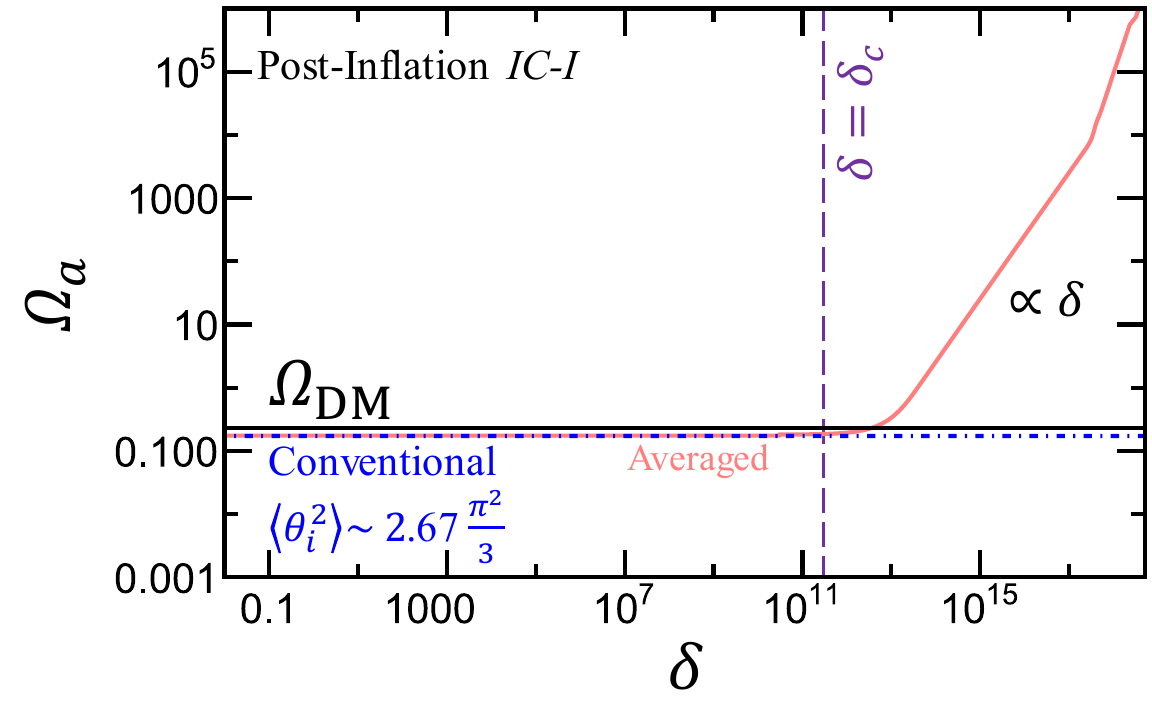}
\caption{\label{Fig:OmegaDelta_IC1_post} The dependence of axion relic abundance on initial velocity ($\delta$) for Type-I IC, post-inflation PQ breaking ($f_a=10^{11}\,$GeV). The kink around $\delta\sim10^{20}$ is due to the change in the number of relativistic degrees of freedom, $g_*$, which is accounted for in our numerical calculation.} 
\end{figure}
\noindent\textbf{\textit{ IC-I: Post-inflationary $U(1)_{PQ}$ breaking.}}\\
In each of the following cases, we consider the field evolution with a random initial field value $\theta_i$ which upon considering $\Omega_a$ will be averaged over post-inflationary causal patches as $\langle \theta_o^2\rangle=\langle \theta_i^2 \rangle \sim 2.67\pi^2/3$ \cite{Visinelli:2009zm,Diez-Tejedor:2017ivd}.\\
{\underline{(i) $0<\delta<\theta_i$}}, i.e., with generally small initial KE. For generic $\theta_i\sim 1$, although the $\frac{R(t)}{R(t_i)}$ term in Eq.~\ref{eq: theta_sol} slightly reduces $\theta(t)$ over time, the effect is negligible relative to the constant term and $\theta_o\sim\theta_i$. Therefore the axion evolution and the prediction $\Omega_a\propto\theta_o^2$ is very similar to the conventional prediction $\Omega_a^{conv}$ and remains so after averaging over causal patches. \\
\noindent{\underline{(ii) $\theta_i<\delta<\delta_c$}}, i.e., with moderate initial KE. In this case KE becomes important at early stages of the field evolution, but is not yet sufficient to cause a notable delay of the $t_o$ relative to $t_o^{con}$. In this regime as $\delta$ increases $\theta_o$ and thus $\Omega_a$ become more sensitive to $\delta$ (oscillatory dependence) due to the periodic nature of the field potential. In particular, $\Omega_a$ can be much suppressed for particular $\delta$ values that causes cancelation among the $\theta_i$ and $\delta$-dependent terms in Eqs.~\ref{eq: sol_IC1}, i.e., when 
\begin{align}
\label{Eq: thetai-delta}
\theta_o = 2\pi k, \;\;\;\;\; k \in \mathds{Z}.
\end{align}
 However, with a constant $\delta$ such an accidental cancelation only occurs for certain $\theta_i$'s, and its effect is washed out after averaging. Consequently, the $\Omega_a$ prediction is comparable to $\Omega_a^{con}$.  \\

\noindent{\underline{(iii) $\delta>\delta_c$}}, i.e., with high initial KE. This case is similar to the above (ii), yet the difference is that KE is reshifted to $\sim V$ after $t_o^{con}$, so the oscillation is delayed. Although $\rho(t_o)$ is the same as in case (ii) at the onset of axion oscillation, the entropy then is diluted as $s(t_o)\sim s(t^{con}_o)\left(t_o^{con}/t_o\right)^{3/2}\propto \delta^{-1}$ (using Eqs.~\ref{eq: sol_IC1}, \ref{eq: tosc}). Like in (ii), accidental cancelation Eq.~\ref{Eq: thetai-delta} can happen but becomes irrelevant after averaging. Therefore $\Omega_a$ is enhanced relative to $\Omega_a^{con}$ by a factor of $\sim\delta/\delta_c$.\\

\noindent{\underline{(iv) $\delta=\theta_i$ or $\delta\approx\theta_i$,}} the special regime where $\delta$ is equal to or in the close vicinity of $\theta_i$. This case is qualitatively different from the previous ones. Due to the (almost) cancelation between $\theta_i$ and $\delta$, $\delta \left( \frac{R(t_i)}{R(t)}\right)$ term in Eq.~\ref{eq: sol_IC1} dominates the evolution which causes a potentially dramatic decrease in $\theta(t)$ until oscillation starts around $t_o\sim t_o^{con}$, with $\theta(t_o)=\theta_o\ll\theta_i$. $\rho_a(t_o)$ would be expected to be suppressed by $\sim\theta_o^2/\theta_i^2$ relative to the conventional. However, he cancelation between $\delta$ and $\theta_i$ only occurs for patches with peculiar $\theta_i$, and the effect disappears after averaging over post-inflationary patches.

\begin{figure}[h] 
\includegraphics[width=0.45\textwidth]{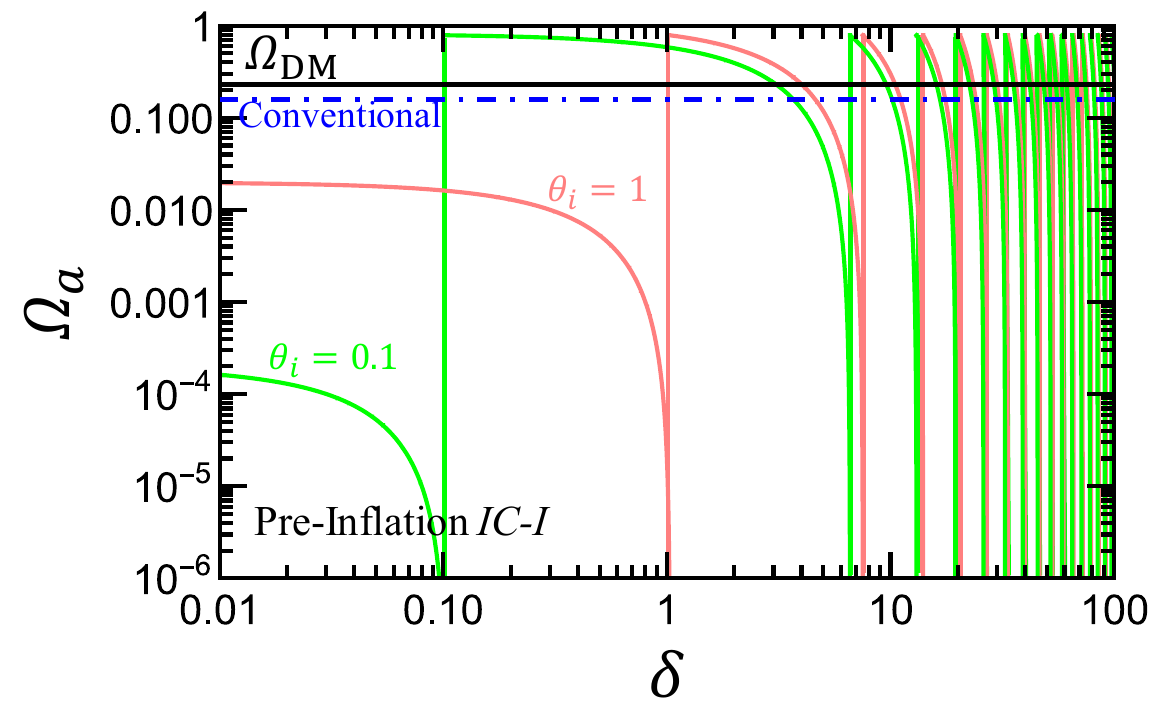}
\caption{ \label{Fig:OmegaDelta_IC1_pre}The dependence of axion relic abundance on initial velocity ($\delta$) for Type-I IC, pre/during-inflation PQ breaking ($f_a=10^{11}\,$GeV, $t_I = 10^3 t_i$). The oscillatory dependence in large $\delta$ region is shown.}
\end{figure}

Detailed illustrations for the field evolution in each of these cases can be found in Fig.~S1 in the Appendix.~A. A representative example of $\Omega_a$-$\delta$ relation is shown in Fig.~\ref{Fig:OmegaDelta_IC1_post} based on numerical results. The analytical estimate of $\Omega_a$ is summarized later in Eq.~\ref{Eq: Omega_a_1}.
\textit{The upshot for the post-inflation scenario is: $\Omega_a$ can be similar to or enhanced relative to the conventional misalignment due to $\dot{\theta}(t_i)\neq0$, while the potential suppression effect in specific patches is resolved after averaging over post-inflationary causal patches.}\\
\noindent\textbf{\textit{{IC-I: Before or during inflation $U(1)_{PQ}$ breaking.}}} During inflation the KE in the axion field is rapidly diluted, while $\theta_i$ or the potential energy freezes in as its value at the onset of inflation (or at PQ breaking time if PQ breaks during inflation). Many of the discussions for the post-inflationary scenario apply here, but there are key differences due to inflationary effects. We briefly summarize the results for the same four cases as follows:\\
{\underline{(i) $0<\delta<\theta_i$:}} similar to the conventional pre-inflationary case.\\
{\underline{(ii) $\theta_i<\delta<\delta_c$:}} in general similar to the conventional, but unlike in the post-inflationary scenario, the accidental cancelation/suppression on $\Omega_a$ (i.e. $\theta_o\approx 0$ in Eq.(\ref{Eq: thetai-delta})) persists without the averaging. \\
{\underline{(iii) $\delta>\delta_c$:}} the situation with $\rho_a$ evolution is similar to the above (iii), but the enhancement due to the diluted $s(t_o)$ is absent due to the intervention of inflation which cuts short the KE domination time (unless inflation happens after QCD phase transition). Therefore the result is in general similar to the conventional, but accidental suppression is possible for certain $\delta$.\\
{\underline{(iv) $\delta=\theta_i$ or $\delta\approx\theta_i$.}} For PQ breaking during inflation, the situation is similar to the conventional, since the initial KE is quickly depleted by inflation before it can drive down $\theta_i$ value. However, if the PQ breaks before inflation with a moderate/large separation in their scales, the $\dot{\theta}(t_i)\neq0$ initial condition can leave a trace despite inflation: the field value is already driven down to $\theta_I \equiv\theta(t_I)\approx\frac{R(t_i)}{R(t_I)}\theta_i$ for $\theta_i-\delta\rightarrow0$, where $t_I$ is when inflation starts.   $\theta(t_I)$ then freezes in as the effective new initial condition for axion misalignment after inflation. \\
\indent  Detailed illustrations for the field evolution in each of these cases can be found in Fig.~S2 in Appendix.~A. A representative example of $\Omega_a$-$\delta$ relation for this scenario is illustrated in Fig.~\ref{Fig:OmegaDelta_IC1_pre} based on numerical results. \textit{The upshot for this pre/during inflation scenario is: $\Omega_a$ is similar to or suppressed relative to the conventional due to $\dot{\theta}(t_i)\neq0$ , while the potential enhancement effective in post-inflationary case is wiped out by inflation.} 

We now summarize the prediction for $\Omega_a$ with IC-I for post- and pre-inflation cases in order:
\begin{align}
\label{Eq: Omega_a_1}
\Omega_a^{\rm post-I} \simeq \left\{            
\begin{aligned}
&\,\Omega_a^{con} =0.02\, \langle\theta_o^2\rangle\left( \frac{f_a}{10^{11}\,\hbox{GeV}} \right)^{7/6}, \hspace{0.1cm} \delta < \delta_c\\*[-0.1\normalbaselineskip]
&\, \Omega_a^{con} \frac{\delta}{\delta_c} \simeq0.01\,  \langle\theta_o^2\rangle \left( \frac{\delta}{10^{11}} \right),\hspace{0.9cm}  \delta \geq \delta_c,
\end{aligned}
\right.
\end{align}
and
\begin{align}\label{Eq: Omega_a_2}
\Omega_a^{\rm pre-I} = \,\Omega_a^{con} \frac{\theta_I^2}{\langle\theta_o^2\rangle},
\end{align}
For the last equality in the 2nd line of Eq.~\ref{Eq: Omega_a_1}, we used Eq.~\ref{Eq: delta_c} which assumes $t_i$ at PQ breaking scale $t_i \sim m_{pl}/f_a^2$. Other $t_i$ choices would change the numerics. $\theta_I$ in Eq.~\ref{Eq: Omega_a_2} is obtained from Eq.~\ref{eq: sol_IC1} with $t=t_I$, which as discussed can lead to a suppression in $\Omega_a^{\rm pre-I}$ when $\theta_i\approx\delta$.
\\
\begin{figure}[h] 
\includegraphics[width=0.46\textwidth]{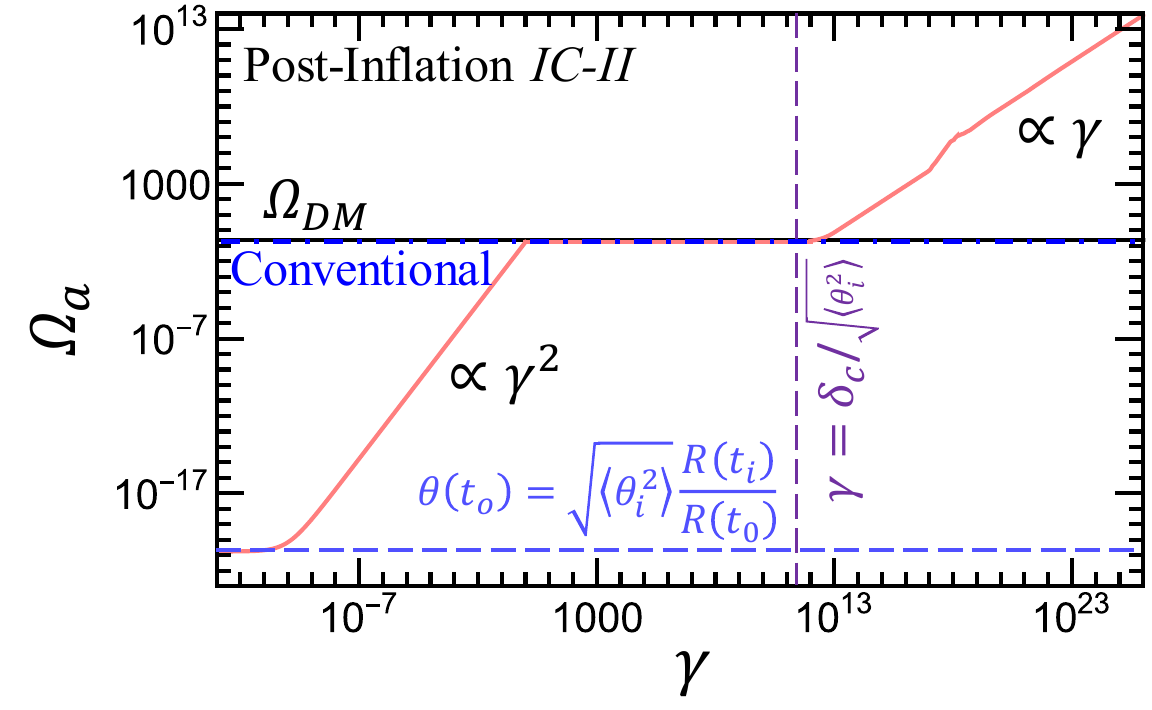}
\caption{\label{Fig: Omega_IC2_postI}. The dependence of axion relic abundance on initial velocity ($\gamma$) for Type-II IC ($f_a=10^{11}$ GeV), post-inflationary PQ-breaking. }
\end{figure}
\indent Note that \textit{case (iv) of this scenario provides a new possibility for $\Omega_a$ to account for $\Omega_{DM}$ with $f_a\gtrsim10^{11}$ GeV} due to specific relations between $\theta_i$ and $\dot{\theta}_i$ even with a natural $\theta_i\sim1$. On the other hand, one could argue that this is trading one type of fine-tuning for another type. Although challenging it is curious to see if it is possible to distinguish the two types of fine-tuning combining various avenues of observational data. \\

\begin{figure}[t] 
\includegraphics[width=0.45\textwidth]{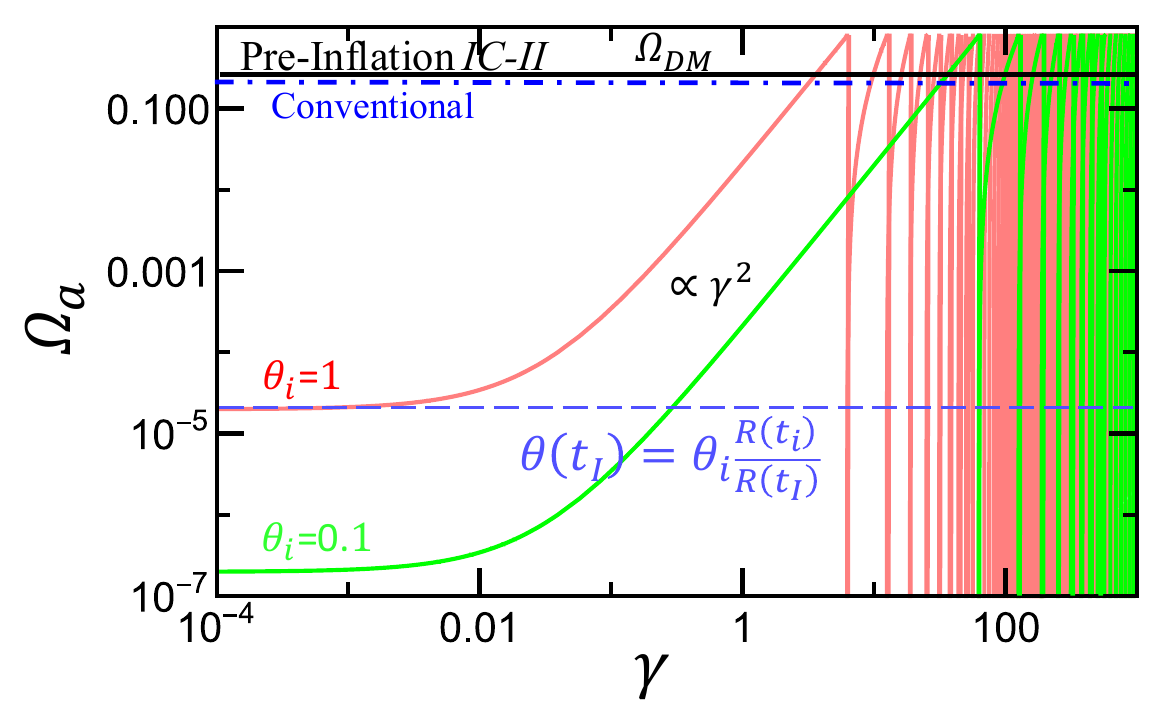}
\caption{\label{Fig: Omega_IC2_preI}. The dependence of axion relic abundance on initial velocity ($\gamma$) for Type-II IC ($f_a=10^{11}$ GeV and $t_I = 10^3 t_i$), pre/during-inflationary PQ-breaking. }
\end{figure}

\textbf{\underline{Type-II IC: $\dot{\theta}_i=-(1-\gamma)\theta_iH_i$}}, where $\gamma \geq 0$. This IC is inspired by case (iv) with Type-I IC. Most results in IC-I apply, with the replacement of $\delta\rightarrow (1-\gamma)\theta_i$ in Eq.~\ref{eq: sol_IC1}. The key difference is that here the relation $\dot{\theta}_i\propto\theta_i$ is assumed to be valid for any $\theta_i$, \textit{therefore the aforementioned suppression effect in case (iv) (here $\gamma\rightarrow 0$) is robust and survives even after averaging patches for post-inflationary PQ breaking.} The realization of such an IC generally requires an explicit PQ breaking term in the axion potential effective in the early Universe, which is beyond the scope of this work but would be interesting to explore. \textit{We highlight this possibility since it provides a novel dynamic way to accommodate large $f_a\gtrsim10^{11}$ GeV QCD axion as a DM candidate for post-inflation PQ breaking.} Despite requiring a special relation between $\theta_i$ and $\dot{\theta}_i$, this solution is intriguing considering that for post-inflationary PQ breaking in the conventional misalignment there is no way to even fine-tune to accommodate $f_a\gtrsim10^{11}$ GeV, since $\theta_i$ is averaged to $O(1)$.  We show the $\Omega_a$-$\gamma$ relation for IC-II in Figs.~\ref{Fig: Omega_IC2_postI} and \ref{Fig: Omega_IC2_preI}, for post- and pre-inflation PQ-breaking, respectively.

\indent The main difference from IC-I in the $\Omega_a$ prediction, i.e., for case (iv) (small $\gamma$) in the post-inflation scenario, is demonstrated with the following formula:
\begin{align}\label{Eq: O-IC2}
\Omega_a^{\rm post-I} \simeq \left\{            
\begin{aligned}& \,\Omega^{con}_a \gamma^2, \hspace{1.3cm} \frac{R(t_i)}{R(t_{o})}  \leq \gamma \leq 1\\
&\, \Omega^{con}_a  \left(\frac{R(t_i)}{R(t_{o})} \right)^{2}  , \hspace{0.45cm} 0 \leq \gamma \leq \frac{R(t_i)}{R(t_{o})},  
\end{aligned}
\right.
\end{align}
where we can clearly see the suppression relative to the conventional by a factor of $\gamma^2$ or $\left( \frac{R(t_i)}{R(t_{o})}\right)^2$.

\section{Conclusion}
In this work, we propose a new misalignment mechanism where the axion initial velocity is non-zero and demonstrate its impact on axion relic abundance based on systematically classified benchmark cases with a UV model-independent approach. While in many cases $\Omega_a$ remains similar to the conventional prediction, it may be significantly enhanced with a large initial velocity $\dot{\theta}_i$ or suppressed when $\dot{\theta}_i$ and $\theta_i$ satisfy peculiar relations. 
As an outcome of this new scenario, new viable parameter space for the QCD axion DM opens up, allowing $f_a$ both much above (in the special suppression region identified in IC-II) \textit{and} much below (with sufficiently large $\dot{\theta}_i$) the conventional $\sim10^{11}$ GeV scale (detailed relations given in Eqs.~\ref{Eq: Omega_a_1}-\ref{Eq: O-IC2}). Detailed realization of these initial conditions and phenomenological consequences for both QCD axion and general ALPs are worth further investigation (an example is given in Appendix.~B). Meanwhile a caveat for the scenario of post-inflationary PQ-breaking is that, as in conventional misalignment, topological defects' contribution to $\Omega_a$ may dominate over the misalignment contribution, and is worth revisiting to advance and complete our understanding of the axion production mechanism \cite{Kawasaki:2014sqa, Chang:2019mza, furture, Gorghetto:2018myk, Buschmann:2019icd}. \\
\noindent\textit{{Side Notes: During the final stage of completing this manuscript, \cite{Co:2019jts} (as also noted in the introduction) came out which has overlap with our results. The two papers are complementary to each other as explained in the introduction.}}

\begin{acknowledgments}
We thank Anson Hook for discussion. The authors are supported in part by the US Department of Energy grant DE-SC0008541. CC thanks Yin Chin Foundation of U.S.A. for its support. YC thanks the Kavli Institute for Theoretical Physics (supported by the National Science Foundation under Grant No. NSF PHY-1748958), Erwin Schr\"odinger International Institute, and Galileo Galilei Institute for hospitality while the work was being completed.
\end{acknowledgments}

\appendix

\section{The dynamics of Axion Evolution with An Initial Velocity}

\begin{figure}[H]\begin{center}
\includegraphics[width=0.478\textwidth]{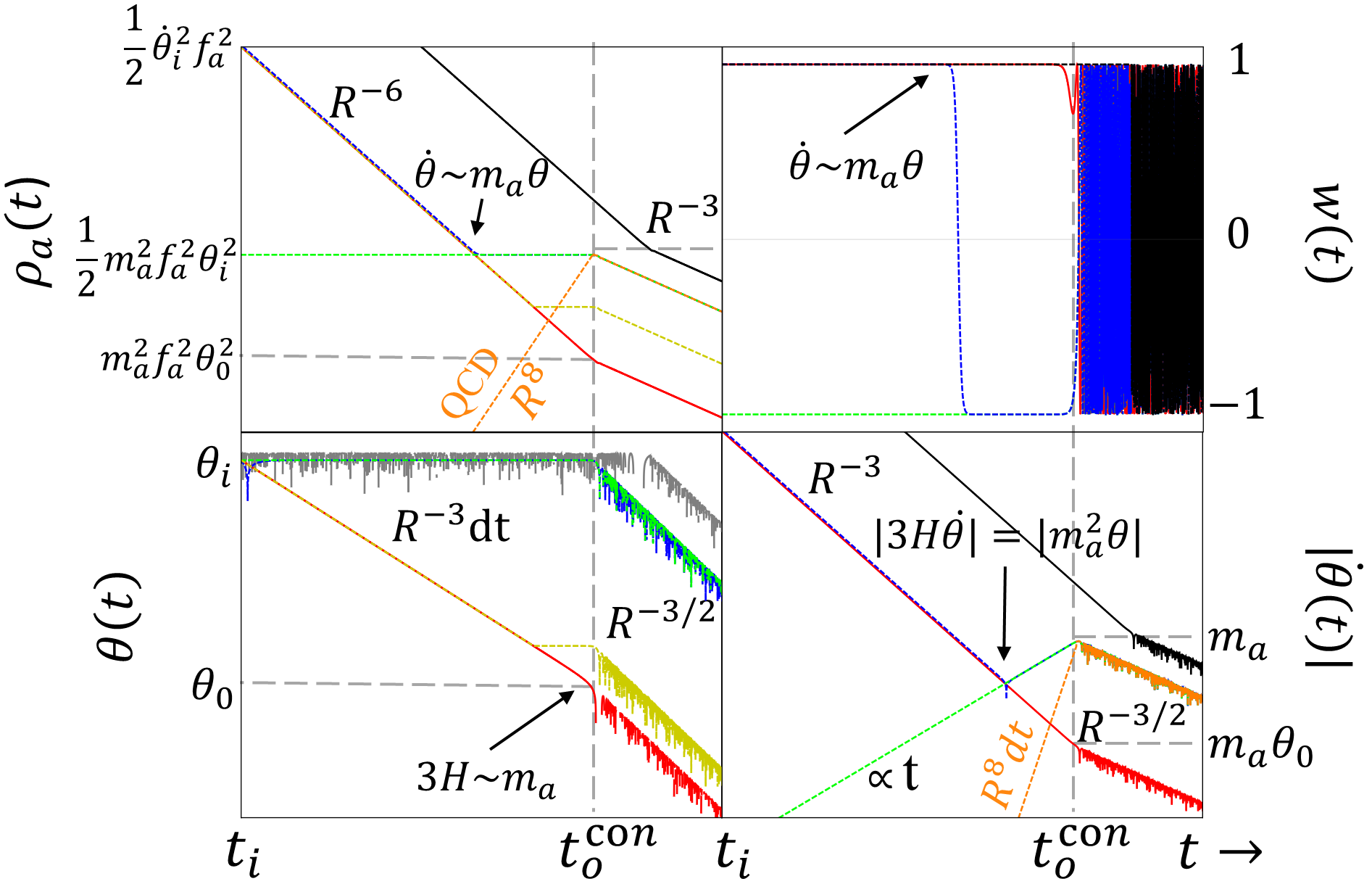}\end{center}
\caption{\label{postInf_evol_S} The time evolution of the axion field with Type-I IC in post-inflation PQ-breaking scenario (for an individual $\theta_i$, before averaging). Green/orange: case (i), blue: case (ii), black: case (iii), red: case (iv). In addition, we assume the axion mass $m_a^2 \propto R^{8}$ as QCD axion in orange curve, others are assumed as constant. Details are also given in the text.}
\end{figure}

\begin{figure}[H]\begin{center}
\includegraphics[width=0.478\textwidth]{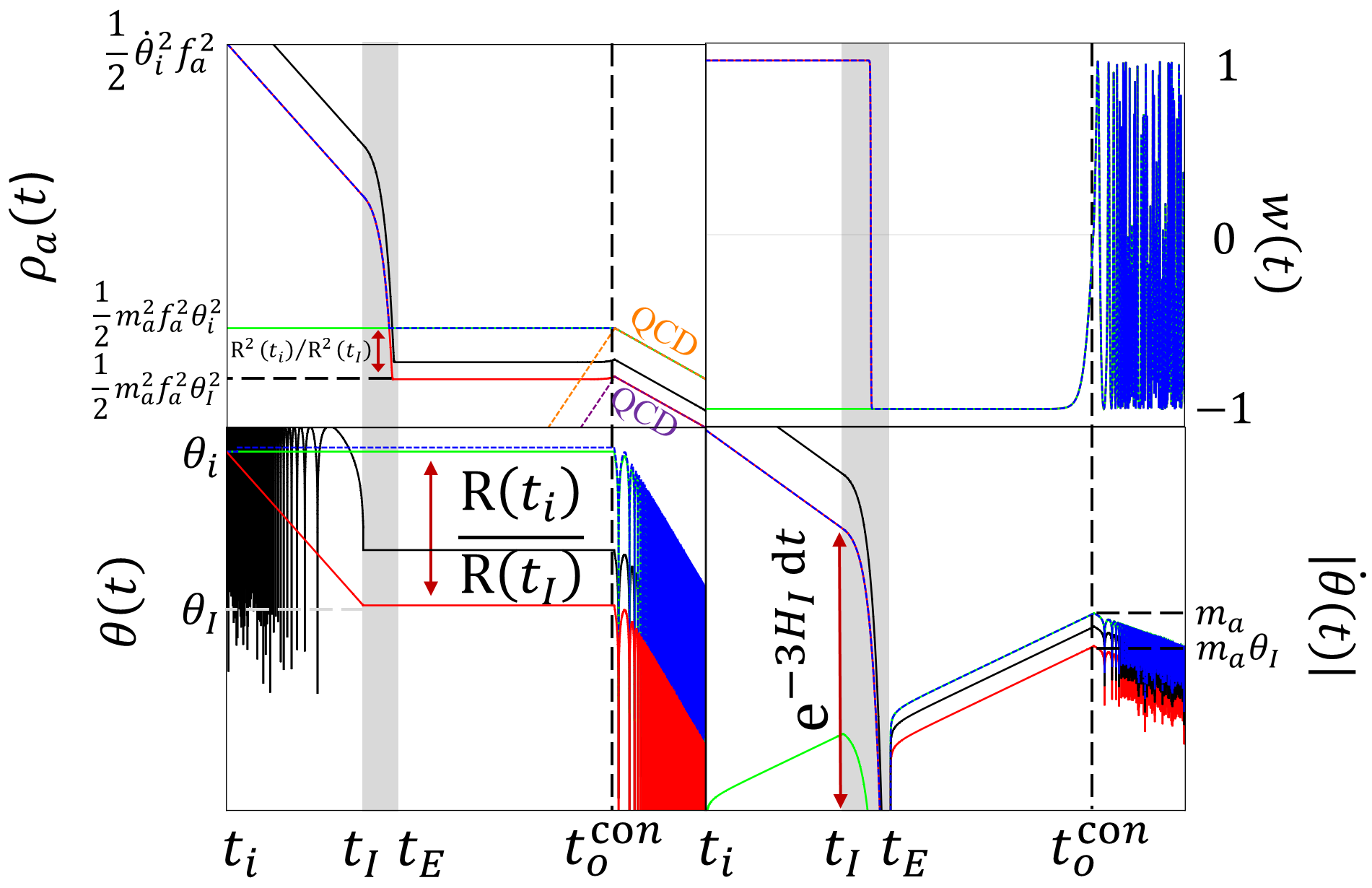}\end{center}
\caption{\label{preInf_evol_S} The time evolution of the axion field with with Type-I IC, in the pre-inflationary PQ-breaking scenario. Color codes are the same as in Fig.~\ref{postInf_evol_S}. In addition, we assume the axion mass $m_a^2 \propto R^{8}$ as QCD axion in orange and purple curves, others are assumed as constant. Details given in the text.} 
\end{figure}

In this section, we illustrate the time evolution of $\theta(t), |\dot{\theta}(t)|, \rho_a(t)$ for the different scenarios of non-zero $\dot{\theta}_i$ in the main text. We will refer to Type-I IC to be specific, while each of the cases may apply to Type-II IC by the substitution of $\delta\rightarrow (1-\gamma)\theta_i$ in $\theta(t)$ solution as \begin{align}\label{eq: sol_IC1}
\theta(t)=\theta_i \left[ \gamma +(1-\gamma) \left( \frac{R(t_i)}{R(t)}\right) \right],~~{\rm Mod}[2\pi].
\end{align} The difference between IC-I and IC-II in predicting $\Omega_a$ is explained in the main text. The results shown in Figs.~\ref{postInf_evol_S}, \ref{preInf_evol_S} are obtained by solving the equation of motion \begin{align}
\label{Eq:EoM}
\ddot{\theta} + 3 H \dot{\theta} + m_a^2(T) \theta=0. 
\end{align}\\

By solving the above E.O.M, we can also obtain the averaged background energy density $\bar{\rho}_a$ and pressure $\bar{P}_a$ of the axion field are:
\begin{align}
\bar{\rho}_a = \frac{1}{2} \dot{a}^2 + \frac{1}{2} m_a^2 a^2, \;\;\;\;\;\;
\bar{P}_a = \frac{1}{2} \dot{a}^2 - \frac{1}{2} m_a^2 a^2.
\end{align}
The equation of state can then be found by applying
\begin{align}
w(t) \equiv \frac{\bar{P}_a}{\bar{\rho}_a}.
\end{align}

\textit{\textbf{IC-I Post-inflationary $U(1)_{PQ}$ breaking:}} the representative solutions for this scenario in the following four cases are illustrated in Fig.~\ref{postInf_evol_S}. $\theta_i$ takes a random initial value which will be averaged over post-inflationary patches for $\Omega_a$ calculation\\
{\underline{(i) $0<\delta<\theta_i$}}, i.e., with generally small initial KE. This is illustrated with green dashed lines ($\delta \sim 0$, overlapping with conventional case) in Fig.~\ref{postInf_evol_S} for constant $m_a$, while for QCD axion the evolution before $t_{o}^{con}$ follows the orange line (overlapping with green for $w(t)$ and $\theta(t)$).

\noindent{\underline{(ii) $\theta_i<\delta<\delta_c$}}, i.e., with moderate initial KE. The evolution for this case is illustrated with blue lines  in Fig.~\ref{postInf_evol_S} (for $\theta_i$ away from the cancelation region $\theta_o = 2\pi k, \;\; k \in \mathds{Z}$). Due to the dominance of KE over $V$ in early times, both the constant $m_a$ and the QCD cases essentially follow same evolution (this also applies to the case (iii) and (iv) below). We can see that due to the large $\theta_i$ the field's equation of state in early stage is kination-like.

\noindent{\underline{(iii) $\delta>\delta_c$}}, i.e., with high initial KE.  This case is illustrated with black lines in Fig.~\ref{postInf_evol_S}.

\noindent{\underline{(iv) $\delta=\theta_i$ or $\delta\approx\theta_i$,}} the special regime where $\delta$ is equal to or in the close vicinity of $\theta_i$. 
This case is demonstrated in red in Fig.~\ref{postInf_evol_S} for $\delta=\theta_i$, in yellow for $\delta\approx\theta_i$.
\textit{\textbf{IC-I Pre-inflationary $U(1)_{PQ}$ breaking:}} the representative solutions for this scenario in the following four cases are illustrated in Fig.~~\ref{postInf_evol_S}, with the same color codes as in their post-inflationary counterparts.
 
\section{An Example Model Generating An Axion Initial Velocity}
In this section, we demonstrate a simple example of generating a nonzero initial velocity of axion starting from $\dot{\theta}(t \to 0) = 0 $, as a result of a breaking of axion shift symmetry in the early Universe. Such a symmetry breaking is analogous to that realized at late times by the QCD instanton effect. This is consistent with the expectation that PQ symmetry, as a global symmetry, is generally considered approximate/accidental. In particular we consider the following effective action involving higher dimensional operators (dimension $\lambda+1$, $\lambda>0$, examples for $V(\Phi_2)$ given in Eq.~\ref{Eq: Vphi2}):
\begin{align}
\notag S  \,\supset - \int dx^4 \sqrt{-g} \Bigg[ \partial^\mu  \Phi^\dagger_i \partial_{\mu} & \Phi_i + \lambda_{1} \left( \Phi^\dagger_1 \Phi_1 - \frac{f_a^2}{2}\right)^2 \\ \notag  & +  g_f \left( \frac{\Phi_2}{\Lambda_2} \right)^\lambda  \Phi_1 + h.c.\Bigg],
\end{align} 
with
\begin{align}
\Phi_1 = \frac{1}{\sqrt{2}} \left( f_a +  \phi  \right)e^{i a/f_a},
\end{align}
where $i=\{1,2\}$, and $\Phi_2, \phi$ are CP-even real scalar. The angular mode of $\Phi_1$ is identified as the axion. For simplicity, we assume the effective coupling $g_f$ is real. The $g_f$ term explicitly breaks the axion shift symmetry, considering the motivated possibility that $\Phi_2$ may be displaced from its true vacuum ($\Phi_2=0$) at the end of inflation. $\Lambda_2$ is the cutoff scale of this effective action, and $\Lambda_2 \gg f_a$. The effective potential of the axion then takes the following form:
\begin{align}
\label{Eq: Lang}
\mathcal{L} \supset - \sqrt{2} g_f f_a  \left(\frac{\Phi_2}{\Lambda_2}\right)^\lambda \cos\left( \frac{a}{f_a} + \alpha_0 \right). 
\end{align}
With a nonzero initial field value, the dynamics of the field $\Phi_2$ provides a variety of possibilities to generate a nontrivial axion velocity at $t_{\hbox{\st{osc}}}$ (the onset of axion oscillation). For instance, we consider that $V(\Phi_2)$ takes the following power-law form, which can occur in e.g. quintessence models \cite{Salati:2002md,Liddle:1998xm,Chung:2007vz,Poulin:2018dzj}:
\begin{align}\label{Eq: Vphi2}
V(\Phi_2)  =g_N\Phi_2^N, \;\;\;\;\; 
\end{align}
where $g_N$ is a constant parameter. We simply assume that this term dominates the potential in early universe. With the dominant background energy density generically parametrized as $\rho \propto R^{-m}$, we find $\rho_{\Phi_2} \propto R^{-n}$, where $n= \left(\frac{N}{N-2}\right)m$.
By solving the time evolution of $\Phi_2$ in this potential, we find
\begin{align}
\label{Eq: supp: condition-0}
\Phi_2(t) \propto t^{1-N/(N-2)}, \;\;\; \forall \, N>2., \;\;n\neq m, \;\; \frac{6-n}{m} >0,
\end{align}
Plugging this solution to Eq.(\ref{Eq: Lang}), we obtain the axion potential at early times:
\begin{align}
\label{Eq: Potential}
V_\Lambda(a) = \Lambda^4 \left( \frac{t_2}{t}\right)^{p} \cos\left( \frac{a}{f_a} + \alpha_0 \right),
\end{align}
where $\Lambda^4 \equiv  \sqrt{2} g_f f_a (M_2/\Lambda_2)^\lambda$ with initial field value of $\Phi_2 (t_2) = M_2$, $p \equiv -\lambda(1-N/(N-2))>0$. With the $V_\Lambda (a)$ given above, we can write down the parametrized equation of motion for the axion field:
\begin{align}
\label{Eq: Supp EoM}
\notag  \ddot{a} + \frac{6}{m} \frac{1}{t} \dot{a} & \, -  \frac{\Lambda^4}{f_a} \left(\frac{t_2}{t}\right)^p  \sin \left(\frac{a}{f_a} + \alpha_0 \right) \\ & \, + m_a^2(T) f_a \sin \left(\frac{a}{f_a} \right) = 0.
\end{align}
A key feature of this model with positive $p$ is that the effect of the potential $V_\Lambda(a)$ is suppressed/negligible after $t\sim 1/m_a$. Consequently, the conventional axion dynamics is restored after QCD phase transition. Therefore the axion has enough time to enter into QCD vacuum and behaves just like the standard QCD axion at late times. We also find that the constraint of solving strong CP problem is weaker than the DM relic density constraint. To see this, we know the following constraint from neutron EDM measurements:
\begin{align}
\theta(t_0) < \theta_{QCD} \simeq 10^{-10}. \label{eq:strongCP}
\end{align}
We also know that axion energy density today is
\begin{align*}
\rho_a (t_0) = m_a^2 f_a^2 \langle\theta^2(t_0)\rangle = 3.49 \times 10^{-5}\,\hbox{GeV}^{2} \langle\theta^2(t_0)\rangle,
\end{align*}
where we substituted $m_a f_a = 5.7\times 10^{-3} \,$GeV$^2$ \cite{diCortona:2015ldu}.
Therefore Eq.~\ref{eq:strongCP} implies an upper bound on axion relic abundance today. Putting all numerics together, we find
$\Omega_a \lesssim 10^{22}$, which is a much more forgiving bound than the DM over-closure bound. Therefore the strong CP constraint can be satisfied as long as $\Omega_{a}\leq\Omega_{\rm DM}$, and the implied constraints on model parameters can be found in Eqs.~10-12 in the main text. 

The example we demonstrated provides a class of models parametrized by $p$, $\lambda$ and $m$, that can generate diverse possibilities of initial conditions for the axion field including a sizable initial velocity. In the following, we will show a choice of parameters that can generate our IC-I (iv) and IC-II.
$\;$\\

\subsubsection{Example: $p=2$, $\lambda=10$, $m=4$}
Here we consider $p=2$, $\lambda=10$, $m=4$, consequently $N=12$ and $n=4.8$. In this case, the equation of motion of axion in the very early Universe can be approximately as the following (the choice of $\alpha_0 = \pi$ leads to $c>0$, while $\alpha_0=0$ leads to $c<0$, and QCD axion potential is negligible during this early era of interest):
\begin{align}
\ddot{a} + \frac{3}{t} \dot{a} + \frac{c}{t^2} a = 0.
\end{align}
where define $|c| \equiv \Lambda^4t_2^2/f_a^2$. The solution is 
\begin{align}
\label{Eq:at}
a(t) = c_1 t^{n_+} + c_2 t^{n_-},
\end{align}
where
\begin{align}
\label{Eq:n pm}
n_{\pm}= \frac{1}{2}\left(-2 \pm \sqrt{4-4c} \right).
\end{align}
In general, $c$ is suppressed by PQ-scale $f_a$ and high energy physics scale $\Lambda_2$. Therefore, we expect $-1 \leq c \leq 1$ and consequently $-1 \leq n_+ \lesssim 0.414$. Assuming that initially at $t_2$, $a(t_2) \sim \pi f_a$ and $\dot{a}(t_2)=0$, we find that $n_+$ term will dominate at late times, and the axion follows the relation
\begin{align}
\label{Eq:n to t}
\frac{\dot{a}(t)}{a(t)} \simeq n_+ \frac{1}{t} = n_+ \frac{m}{2} H(t). 
\end{align}
To give a numerical benchmark example, we consider
\begin{align*}
& \,M_2 \sim 10^5 \hbox{GeV}, \;\;\; g_N \sim 1\, \hbox{GeV}^{-8}, \;\;\; \\ & \, \Lambda \sim 10 \sqrt{m_a f_a}, \;\;\; m_a \sim 10^{-15}\,\hbox{GeV}, \;\;\; \\ & \, t_2 \sim  10^{-2} t_o^{con}, \;\;\;\alpha_0=\pi.
\end{align*}
This implies
\begin{align}
c = \Lambda^4 \frac{t_2^2}{f_a^2} \sim 1 \;\;\; \to \;\;\; n_+ \sim -1, 
\end{align}
therefore
\begin{align}
\frac{\dot{a}(t_o^{con})}{a(t_o^{con})} = \frac{\dot{\theta}(t_o^{con})}{\theta(t_o^{con})} \sim n_+ \frac{m}{2} H(t_o^{con}) \sim - H(t_o^{con}).
\end{align}
This gives our IC-I-(\textit{iv}) and IC-II.

\bibliographystyle{apsrev4-1}
\bibliography{REsub_to_ArXiv}

\begin{thebibliography}{45}%
\makeatletter
\providecommand \@ifxundefined [1]{%
 \@ifx{#1\undefined}
}%
\providecommand \@ifnum [1]{%
 \ifnum #1\expandafter \@firstoftwo
 \else \expandafter \@secondoftwo
 \fi
}%
\providecommand \@ifx [1]{%
 \ifx #1\expandafter \@firstoftwo
 \else \expandafter \@secondoftwo
 \fi
}%
\providecommand \natexlab [1]{#1}%
\providecommand \enquote  [1]{``#1''}%
\providecommand \bibnamefont  [1]{#1}%
\providecommand \bibfnamefont [1]{#1}%
\providecommand \citenamefont [1]{#1}%
\providecommand \href@noop [0]{\@secondoftwo}%
\providecommand \href [0]{\begingroup \@sanitize@url \@href}%
\providecommand \@href[1]{\@@startlink{#1}\@@href}%
\providecommand \@@href[1]{\endgroup#1\@@endlink}%
\providecommand \@sanitize@url [0]{\catcode `\\12\catcode `\$12\catcode
  `\&12\catcode `\#12\catcode `\^12\catcode `\_12\catcode `\%12\relax}%
\providecommand \@@startlink[1]{}%
\providecommand \@@endlink[0]{}%
\providecommand \url  [0]{\begingroup\@sanitize@url \@url }%
\providecommand \@url [1]{\endgroup\@href {#1}{\urlprefix }}%
\providecommand \urlprefix  [0]{URL }%
\providecommand \Eprint [0]{\href }%
\providecommand \doibase [0]{http://dx.doi.org/}%
\providecommand \selectlanguage [0]{\@gobble}%
\providecommand \bibinfo  [0]{\@secondoftwo}%
\providecommand \bibfield  [0]{\@secondoftwo}%
\providecommand \translation [1]{[#1]}%
\providecommand \BibitemOpen [0]{}%
\providecommand \bibitemStop [0]{}%
\providecommand \bibitemNoStop [0]{.\EOS\space}%
\providecommand \EOS [0]{\spacefactor3000\relax}%
\providecommand \BibitemShut  [1]{\csname bibitem#1\endcsname}%
\let\auto@bib@innerbib\@empty
\bibitem [{\citenamefont {Peccei}\ and\ \citenamefont
  {Quinn}(1977{\natexlab{a}})}]{Peccei:1977hh}%
  \BibitemOpen
  \bibfield  {author} {\bibinfo {author} {\bibfnamefont {R.~D.}\ \bibnamefont
  {Peccei}}\ and\ \bibinfo {author} {\bibfnamefont {H.~R.}\ \bibnamefont
  {Quinn}},\ }\href {\doibase 10.1103/PhysRevLett.38.1440} {\bibfield
  {journal} {\bibinfo  {journal} {Phys. Rev. Lett.}\ }\textbf {\bibinfo
  {volume} {38}},\ \bibinfo {pages} {1440} (\bibinfo {year}
  {1977}{\natexlab{a}})},\ \bibinfo {note} {[,328(1977)]}\BibitemShut {NoStop}%
\bibitem [{\citenamefont {Peccei}\ and\ \citenamefont
  {Quinn}(1977{\natexlab{b}})}]{Peccei:1977ur}%
  \BibitemOpen
  \bibfield  {author} {\bibinfo {author} {\bibfnamefont {R.~D.}\ \bibnamefont
  {Peccei}}\ and\ \bibinfo {author} {\bibfnamefont {H.~R.}\ \bibnamefont
  {Quinn}},\ }\href {\doibase 10.1103/PhysRevD.16.1791} {\bibfield  {journal}
  {\bibinfo  {journal} {Phys. Rev.}\ }\textbf {\bibinfo {volume} {D16}},\
  \bibinfo {pages} {1791} (\bibinfo {year} {1977}{\natexlab{b}})}\BibitemShut
  {NoStop}%
\bibitem [{\citenamefont {Wilczek}(1978)}]{Wilczek:1977pj}%
  \BibitemOpen
  \bibfield  {author} {\bibinfo {author} {\bibfnamefont {F.}~\bibnamefont
  {Wilczek}},\ }\href {\doibase 10.1103/PhysRevLett.40.279} {\bibfield
  {journal} {\bibinfo  {journal} {Phys. Rev. Lett.}\ }\textbf {\bibinfo
  {volume} {40}},\ \bibinfo {pages} {279} (\bibinfo {year} {1978})}\BibitemShut
  {NoStop}%
\bibitem [{\citenamefont {Weinberg}(1978)}]{Weinberg:1977ma}%
  \BibitemOpen
  \bibfield  {author} {\bibinfo {author} {\bibfnamefont {S.}~\bibnamefont
  {Weinberg}},\ }\href {\doibase 10.1103/PhysRevLett.40.223} {\bibfield
  {journal} {\bibinfo  {journal} {Phys. Rev. Lett.}\ }\textbf {\bibinfo
  {volume} {40}},\ \bibinfo {pages} {223} (\bibinfo {year} {1978})}\BibitemShut
  {NoStop}%
\bibitem [{\citenamefont {Abbott}\ and\ \citenamefont
  {Sikivie}(1983)}]{Abbott:1982af}%
  \BibitemOpen
  \bibfield  {author} {\bibinfo {author} {\bibfnamefont {L.~F.}\ \bibnamefont
  {Abbott}}\ and\ \bibinfo {author} {\bibfnamefont {P.}~\bibnamefont
  {Sikivie}},\ }\href {\doibase 10.1016/0370-2693(83)90638-X} {\bibfield
  {journal} {\bibinfo  {journal} {Phys. Lett.}\ }\textbf {\bibinfo {volume}
  {120B}},\ \bibinfo {pages} {133} (\bibinfo {year} {1983})}\BibitemShut
  {NoStop}%
\bibitem [{\citenamefont {Dine}\ and\ \citenamefont
  {Fischler}(1983)}]{Dine:1982ah}%
  \BibitemOpen
  \bibfield  {author} {\bibinfo {author} {\bibfnamefont {M.}~\bibnamefont
  {Dine}}\ and\ \bibinfo {author} {\bibfnamefont {W.}~\bibnamefont
  {Fischler}},\ }\href {\doibase 10.1016/0370-2693(83)90639-1} {\bibfield
  {journal} {\bibinfo  {journal} {Phys. Lett.}\ }\textbf {\bibinfo {volume}
  {120B}},\ \bibinfo {pages} {137} (\bibinfo {year} {1983})}\BibitemShut
  {NoStop}%
\bibitem [{\citenamefont {Preskill}\ \emph {et~al.}(1983)\citenamefont
  {Preskill}, \citenamefont {Wise},\ and\ \citenamefont
  {Wilczek}}]{Preskill:1982cy}%
  \BibitemOpen
  \bibfield  {author} {\bibinfo {author} {\bibfnamefont {J.}~\bibnamefont
  {Preskill}}, \bibinfo {author} {\bibfnamefont {M.~B.}\ \bibnamefont {Wise}},
  \ and\ \bibinfo {author} {\bibfnamefont {F.}~\bibnamefont {Wilczek}},\ }\href
  {\doibase 10.1016/0370-2693(83)90637-8} {\bibfield  {journal} {\bibinfo
  {journal} {Phys. Lett.}\ }\textbf {\bibinfo {volume} {120B}},\ \bibinfo
  {pages} {127} (\bibinfo {year} {1983})}\BibitemShut {NoStop}%
\bibitem [{\citenamefont {Irastorza}\ and\ \citenamefont
  {Redondo}(2018)}]{Irastorza:2018dyq}%
  \BibitemOpen
  \bibfield  {author} {\bibinfo {author} {\bibfnamefont {I.~G.}\ \bibnamefont
  {Irastorza}}\ and\ \bibinfo {author} {\bibfnamefont {J.}~\bibnamefont
  {Redondo}},\ }\href {\doibase 10.1016/j.ppnp.2018.05.003} {\bibfield
  {journal} {\bibinfo  {journal} {Prog. Part. Nucl. Phys.}\ }\textbf {\bibinfo
  {volume} {102}},\ \bibinfo {pages} {89} (\bibinfo {year} {2018})},\ \Eprint
  {http://arxiv.org/abs/1801.08127} {arXiv:1801.08127 [hep-ph]} \BibitemShut
  {NoStop}%
\bibitem [{\citenamefont {Marsh}(2016)}]{Marsh:2015xka}%
  \BibitemOpen
  \bibfield  {author} {\bibinfo {author} {\bibfnamefont {D.~J.~E.}\
  \bibnamefont {Marsh}},\ }\href {\doibase 10.1016/j.physrep.2016.06.005}
  {\bibfield  {journal} {\bibinfo  {journal} {Phys. Rept.}\ }\textbf {\bibinfo
  {volume} {643}},\ \bibinfo {pages} {1} (\bibinfo {year} {2016})},\ \Eprint
  {http://arxiv.org/abs/1510.07633} {arXiv:1510.07633 [astro-ph.CO]}
  \BibitemShut {NoStop}%
\bibitem [{\citenamefont {Sikivie}(1982)}]{Sikivie:1982qv}%
  \BibitemOpen
  \bibfield  {author} {\bibinfo {author} {\bibfnamefont {P.}~\bibnamefont
  {Sikivie}},\ }\href {\doibase 10.1103/PhysRevLett.48.1156} {\bibfield
  {journal} {\bibinfo  {journal} {Phys. Rev. Lett.}\ }\textbf {\bibinfo
  {volume} {48}},\ \bibinfo {pages} {1156} (\bibinfo {year}
  {1982})}\BibitemShut {NoStop}%
\bibitem [{\citenamefont {Vilenkin}(1985)}]{Vilenkin:1984ib}%
  \BibitemOpen
  \bibfield  {author} {\bibinfo {author} {\bibfnamefont {A.}~\bibnamefont
  {Vilenkin}},\ }\href {\doibase 10.1016/0370-1573(85)90033-X} {\bibfield
  {journal} {\bibinfo  {journal} {Phys. Rept.}\ }\textbf {\bibinfo {volume}
  {121}},\ \bibinfo {pages} {263} (\bibinfo {year} {1985})}\BibitemShut
  {NoStop}%
\bibitem [{\citenamefont {Davis}(1986)}]{Davis:1986xc}%
  \BibitemOpen
  \bibfield  {author} {\bibinfo {author} {\bibfnamefont {R.~L.}\ \bibnamefont
  {Davis}},\ }\href {\doibase 10.1016/0370-2693(86)90300-X} {\bibfield
  {journal} {\bibinfo  {journal} {Phys. Lett.}\ }\textbf {\bibinfo {volume}
  {B180}},\ \bibinfo {pages} {225} (\bibinfo {year} {1986})}\BibitemShut
  {NoStop}%
\bibitem [{\citenamefont {Vincent}\ \emph {et~al.}(1997)\citenamefont
  {Vincent}, \citenamefont {Hindmarsh},\ and\ \citenamefont
  {Sakellariadou}}]{Vincent:1996rb}%
  \BibitemOpen
  \bibfield  {author} {\bibinfo {author} {\bibfnamefont {G.~R.}\ \bibnamefont
  {Vincent}}, \bibinfo {author} {\bibfnamefont {M.}~\bibnamefont {Hindmarsh}},
  \ and\ \bibinfo {author} {\bibfnamefont {M.}~\bibnamefont {Sakellariadou}},\
  }\href {\doibase 10.1103/PhysRevD.56.637} {\bibfield  {journal} {\bibinfo
  {journal} {Phys. Rev.}\ }\textbf {\bibinfo {volume} {D56}},\ \bibinfo {pages}
  {637} (\bibinfo {year} {1997})},\ \Eprint
  {http://arxiv.org/abs/astro-ph/9612135} {arXiv:astro-ph/9612135 [astro-ph]}
  \BibitemShut {NoStop}%
\bibitem [{\citenamefont {Kawasaki}\ \emph {et~al.}(2015)\citenamefont
  {Kawasaki}, \citenamefont {Saikawa},\ and\ \citenamefont
  {Sekiguchi}}]{Kawasaki:2014sqa}%
  \BibitemOpen
  \bibfield  {author} {\bibinfo {author} {\bibfnamefont {M.}~\bibnamefont
  {Kawasaki}}, \bibinfo {author} {\bibfnamefont {K.}~\bibnamefont {Saikawa}}, \
  and\ \bibinfo {author} {\bibfnamefont {T.}~\bibnamefont {Sekiguchi}},\ }\href
  {\doibase 10.1103/PhysRevD.91.065014} {\bibfield  {journal} {\bibinfo
  {journal} {Phys. Rev.}\ }\textbf {\bibinfo {volume} {D91}},\ \bibinfo {pages}
  {065014} (\bibinfo {year} {2015})},\ \Eprint {http://arxiv.org/abs/1412.0789}
  {arXiv:1412.0789 [hep-ph]} \BibitemShut {NoStop}%
\bibitem [{\citenamefont {Vilenkin}\ and\ \citenamefont
  {Vachaspati}(1987)}]{Vilenkin:1986ku}%
  \BibitemOpen
  \bibfield  {author} {\bibinfo {author} {\bibfnamefont {A.}~\bibnamefont
  {Vilenkin}}\ and\ \bibinfo {author} {\bibfnamefont {T.}~\bibnamefont
  {Vachaspati}},\ }\href {\doibase 10.1103/PhysRevD.35.1138} {\bibfield
  {journal} {\bibinfo  {journal} {Phys. Rev.}\ }\textbf {\bibinfo {volume}
  {D35}},\ \bibinfo {pages} {1138} (\bibinfo {year} {1987})}\BibitemShut
  {NoStop}%
\bibitem [{\citenamefont {Klaer}\ and\ \citenamefont
  {Moore}(2017)}]{Klaer:2017qhr}%
  \BibitemOpen
  \bibfield  {author} {\bibinfo {author} {\bibfnamefont {V.~B.}\ \bibnamefont
  {Klaer}}\ and\ \bibinfo {author} {\bibfnamefont {G.~D.}\ \bibnamefont
  {Moore}},\ }\href {\doibase 10.1088/1475-7516/2017/10/043} {\bibfield
  {journal} {\bibinfo  {journal} {JCAP}\ }\textbf {\bibinfo {volume} {1710}},\
  \bibinfo {pages} {043} (\bibinfo {year} {2017})},\ \Eprint
  {http://arxiv.org/abs/1707.05566} {arXiv:1707.05566 [hep-ph]} \BibitemShut
  {NoStop}%
\bibitem [{\citenamefont {Gorghetto}\ \emph {et~al.}(2018)\citenamefont
  {Gorghetto}, \citenamefont {Hardy},\ and\ \citenamefont
  {Villadoro}}]{Gorghetto:2018myk}%
  \BibitemOpen
  \bibfield  {author} {\bibinfo {author} {\bibfnamefont {M.}~\bibnamefont
  {Gorghetto}}, \bibinfo {author} {\bibfnamefont {E.}~\bibnamefont {Hardy}}, \
  and\ \bibinfo {author} {\bibfnamefont {G.}~\bibnamefont {Villadoro}},\ }\href
  {\doibase 10.1007/JHEP07(2018)151} {\bibfield  {journal} {\bibinfo  {journal}
  {JHEP}\ }\textbf {\bibinfo {volume} {07}},\ \bibinfo {pages} {151} (\bibinfo
  {year} {2018})},\ \Eprint {http://arxiv.org/abs/1806.04677} {arXiv:1806.04677
  [hep-ph]} \BibitemShut {NoStop}%
\bibitem [{\citenamefont {Kawasaki}\ \emph {et~al.}(2018)\citenamefont
  {Kawasaki}, \citenamefont {Sekiguchi}, \citenamefont {Yamaguchi},\ and\
  \citenamefont {Yokoyama}}]{Kawasaki:2018bzv}%
  \BibitemOpen
  \bibfield  {author} {\bibinfo {author} {\bibfnamefont {M.}~\bibnamefont
  {Kawasaki}}, \bibinfo {author} {\bibfnamefont {T.}~\bibnamefont {Sekiguchi}},
  \bibinfo {author} {\bibfnamefont {M.}~\bibnamefont {Yamaguchi}}, \ and\
  \bibinfo {author} {\bibfnamefont {J.}~\bibnamefont {Yokoyama}},\ }\href
  {\doibase 10.1093/ptep/pty098} {\bibfield  {journal} {\bibinfo  {journal}
  {PTEP}\ }\textbf {\bibinfo {volume} {2018}},\ \bibinfo {pages} {091E01}
  (\bibinfo {year} {2018})},\ \Eprint {http://arxiv.org/abs/1806.05566}
  {arXiv:1806.05566 [hep-ph]} \BibitemShut {NoStop}%
\bibitem [{\citenamefont {Martins}(2019{\natexlab{a}})}]{Martins:2018dqg}%
  \BibitemOpen
  \bibfield  {author} {\bibinfo {author} {\bibfnamefont {C.~J. A.~P.}\
  \bibnamefont {Martins}},\ }\href {\doibase 10.1016/j.physletb.2018.11.031}
  {\bibfield  {journal} {\bibinfo  {journal} {Phys. Lett.}\ }\textbf {\bibinfo
  {volume} {B788}},\ \bibinfo {pages} {147} (\bibinfo {year}
  {2019}{\natexlab{a}})},\ \Eprint {http://arxiv.org/abs/1811.12678}
  {arXiv:1811.12678 [astro-ph.CO]} \BibitemShut {NoStop}%
\bibitem [{\citenamefont {Buschmann}\ \emph {et~al.}(2019)\citenamefont
  {Buschmann}, \citenamefont {Foster},\ and\ \citenamefont
  {Safdi}}]{Buschmann:2019icd}%
  \BibitemOpen
  \bibfield  {author} {\bibinfo {author} {\bibfnamefont {M.}~\bibnamefont
  {Buschmann}}, \bibinfo {author} {\bibfnamefont {J.~W.}\ \bibnamefont
  {Foster}}, \ and\ \bibinfo {author} {\bibfnamefont {B.~R.}\ \bibnamefont
  {Safdi}},\ }\href@noop {} {\  (\bibinfo {year} {2019})},\ \Eprint
  {http://arxiv.org/abs/1906.00967} {arXiv:1906.00967 [astro-ph.CO]}
  \BibitemShut {NoStop}%
\bibitem [{\citenamefont {Hindmarsh}\ \emph {et~al.}(2019)\citenamefont
  {Hindmarsh}, \citenamefont {Lizarraga}, \citenamefont {Lopez-Eiguren},\ and\
  \citenamefont {Urrestilla}}]{Hindmarsh:2019csc}%
  \BibitemOpen
  \bibfield  {author} {\bibinfo {author} {\bibfnamefont {M.}~\bibnamefont
  {Hindmarsh}}, \bibinfo {author} {\bibfnamefont {J.}~\bibnamefont
  {Lizarraga}}, \bibinfo {author} {\bibfnamefont {A.}~\bibnamefont
  {Lopez-Eiguren}}, \ and\ \bibinfo {author} {\bibfnamefont {J.}~\bibnamefont
  {Urrestilla}},\ }\href@noop {} {\  (\bibinfo {year} {2019})},\ \Eprint
  {http://arxiv.org/abs/1908.03522} {arXiv:1908.03522 [astro-ph.CO]}
  \BibitemShut {NoStop}%
\bibitem [{\citenamefont {Martins}(2019{\natexlab{b}})}]{Martins:2018}%
  \BibitemOpen
  \bibfield  {author} {\bibinfo {author} {\bibfnamefont {C.~J. A.~P.}\
  \bibnamefont {Martins}},\ }\href {\doibase 10.1016/j.physletb.2018.11.031}
  {\bibfield  {journal} {\bibinfo  {journal} {Phys. Lett.}\ }\textbf {\bibinfo
  {volume} {B788}},\ \bibinfo {pages} {147} (\bibinfo {year}
  {2019}{\natexlab{b}})},\ \Eprint {http://arxiv.org/abs/1811.12678}
  {arXiv:1811.12678 [astro-ph.CO]} \BibitemShut {NoStop}%
\bibitem [{\citenamefont {Hook}(2019)}]{Hook:2018dlk}%
  \BibitemOpen
  \bibfield  {author} {\bibinfo {author} {\bibfnamefont {A.}~\bibnamefont
  {Hook}},\ }\href@noop {} {\bibfield  {journal} {\bibinfo  {journal} {PoS}\
  }\textbf {\bibinfo {volume} {TASI2018}},\ \bibinfo {pages} {004} (\bibinfo
  {year} {2019})},\ \Eprint {http://arxiv.org/abs/1812.02669} {arXiv:1812.02669
  [hep-ph]} \BibitemShut {NoStop}%
\bibitem [{\citenamefont {Co}\ \emph {et~al.}(2019)\citenamefont {Co},
  \citenamefont {Hall},\ and\ \citenamefont {Harigaya}}]{Co:2019jts}%
  \BibitemOpen
  \bibfield  {author} {\bibinfo {author} {\bibfnamefont {R.~T.}\ \bibnamefont
  {Co}}, \bibinfo {author} {\bibfnamefont {L.~J.}\ \bibnamefont {Hall}}, \ and\
  \bibinfo {author} {\bibfnamefont {K.}~\bibnamefont {Harigaya}},\ }\href@noop
  {} {\  (\bibinfo {year} {2019})},\ \Eprint {http://arxiv.org/abs/1910.14152}
  {arXiv:1910.14152 [hep-ph]} \BibitemShut {NoStop}%
\bibitem [{\citenamefont {Affleck}\ and\ \citenamefont
  {Dine}(1985)}]{Affleck:1984fy}%
  \BibitemOpen
  \bibfield  {author} {\bibinfo {author} {\bibfnamefont {I.}~\bibnamefont
  {Affleck}}\ and\ \bibinfo {author} {\bibfnamefont {M.}~\bibnamefont {Dine}},\
  }\href {\doibase 10.1016/0550-3213(85)90021-5} {\bibfield  {journal}
  {\bibinfo  {journal} {Nucl. Phys.}\ }\textbf {\bibinfo {volume} {B249}},\
  \bibinfo {pages} {361} (\bibinfo {year} {1985})}\BibitemShut {NoStop}%
\bibitem [{\citenamefont {Dine}\ \emph {et~al.}(1995)\citenamefont {Dine},
  \citenamefont {Randall},\ and\ \citenamefont {Thomas}}]{Dine:1995uk}%
  \BibitemOpen
  \bibfield  {author} {\bibinfo {author} {\bibfnamefont {M.}~\bibnamefont
  {Dine}}, \bibinfo {author} {\bibfnamefont {L.}~\bibnamefont {Randall}}, \
  and\ \bibinfo {author} {\bibfnamefont {S.~D.}\ \bibnamefont {Thomas}},\
  }\href {\doibase 10.1103/PhysRevLett.75.398} {\bibfield  {journal} {\bibinfo
  {journal} {Phys. Rev. Lett.}\ }\textbf {\bibinfo {volume} {75}},\ \bibinfo
  {pages} {398} (\bibinfo {year} {1995})},\ \Eprint
  {http://arxiv.org/abs/hep-ph/9503303} {arXiv:hep-ph/9503303 [hep-ph]}
  \BibitemShut {NoStop}%
\bibitem [{\citenamefont {Dine}\ \emph {et~al.}(1996)\citenamefont {Dine},
  \citenamefont {Randall},\ and\ \citenamefont {Thomas}}]{Dine:1995kz}%
  \BibitemOpen
  \bibfield  {author} {\bibinfo {author} {\bibfnamefont {M.}~\bibnamefont
  {Dine}}, \bibinfo {author} {\bibfnamefont {L.}~\bibnamefont {Randall}}, \
  and\ \bibinfo {author} {\bibfnamefont {S.~D.}\ \bibnamefont {Thomas}},\
  }\href {\doibase 10.1016/0550-3213(95)00538-2} {\bibfield  {journal}
  {\bibinfo  {journal} {Nucl. Phys.}\ }\textbf {\bibinfo {volume} {B458}},\
  \bibinfo {pages} {291} (\bibinfo {year} {1996})},\ \Eprint
  {http://arxiv.org/abs/hep-ph/9507453} {arXiv:hep-ph/9507453 [hep-ph]}
  \BibitemShut {NoStop}%
\bibitem [{\citenamefont {Kamionkowski}\ and\ \citenamefont
  {March-Russell}(1992)}]{Kamionkowski:1992mf}%
  \BibitemOpen
  \bibfield  {author} {\bibinfo {author} {\bibfnamefont {M.}~\bibnamefont
  {Kamionkowski}}\ and\ \bibinfo {author} {\bibfnamefont {J.}~\bibnamefont
  {March-Russell}},\ }\href {\doibase 10.1016/0370-2693(92)90492-M} {\bibfield
  {journal} {\bibinfo  {journal} {Phys. Lett.}\ }\textbf {\bibinfo {volume}
  {B282}},\ \bibinfo {pages} {137} (\bibinfo {year} {1992})},\ \Eprint
  {http://arxiv.org/abs/hep-th/9202003} {arXiv:hep-th/9202003 [hep-th]}
  \BibitemShut {NoStop}%
\bibitem [{\citenamefont {Graham}\ \emph {et~al.}(2015)\citenamefont {Graham},
  \citenamefont {Kaplan},\ and\ \citenamefont {Rajendran}}]{Graham:2015cka}%
  \BibitemOpen
  \bibfield  {author} {\bibinfo {author} {\bibfnamefont {P.~W.}\ \bibnamefont
  {Graham}}, \bibinfo {author} {\bibfnamefont {D.~E.}\ \bibnamefont {Kaplan}},
  \ and\ \bibinfo {author} {\bibfnamefont {S.}~\bibnamefont {Rajendran}},\
  }\href {\doibase 10.1103/PhysRevLett.115.221801} {\bibfield  {journal}
  {\bibinfo  {journal} {Phys. Rev. Lett.}\ }\textbf {\bibinfo {volume} {115}},\
  \bibinfo {pages} {221801} (\bibinfo {year} {2015})},\ \Eprint
  {http://arxiv.org/abs/1504.07551} {arXiv:1504.07551 [hep-ph]} \BibitemShut
  {NoStop}%
\bibitem [{\citenamefont {Grilli~di Cortona}\ \emph {et~al.}(2016)\citenamefont
  {Grilli~di Cortona}, \citenamefont {Hardy}, \citenamefont {Pardo~Vega},\ and\
  \citenamefont {Villadoro}}]{diCortona:2015ldu}%
  \BibitemOpen
  \bibfield  {author} {\bibinfo {author} {\bibfnamefont {G.}~\bibnamefont
  {Grilli~di Cortona}}, \bibinfo {author} {\bibfnamefont {E.}~\bibnamefont
  {Hardy}}, \bibinfo {author} {\bibfnamefont {J.}~\bibnamefont {Pardo~Vega}}, \
  and\ \bibinfo {author} {\bibfnamefont {G.}~\bibnamefont {Villadoro}},\ }\href
  {\doibase 10.1007/JHEP01(2016)034} {\bibfield  {journal} {\bibinfo  {journal}
  {JHEP}\ }\textbf {\bibinfo {volume} {01}},\ \bibinfo {pages} {034} (\bibinfo
  {year} {2016})},\ \Eprint {http://arxiv.org/abs/1511.02867} {arXiv:1511.02867
  [hep-ph]} \BibitemShut {NoStop}%
\bibitem [{\citenamefont {Bonati}\ \emph {et~al.}(2018)\citenamefont {Bonati},
  \citenamefont {D'Elia}, \citenamefont {Martinelli}, \citenamefont {Negro},
  \citenamefont {Sanfilippo},\ and\ \citenamefont {Todaro}}]{Bonati:2018blm}%
  \BibitemOpen
  \bibfield  {author} {\bibinfo {author} {\bibfnamefont {C.}~\bibnamefont
  {Bonati}}, \bibinfo {author} {\bibfnamefont {M.}~\bibnamefont {D'Elia}},
  \bibinfo {author} {\bibfnamefont {G.}~\bibnamefont {Martinelli}}, \bibinfo
  {author} {\bibfnamefont {F.}~\bibnamefont {Negro}}, \bibinfo {author}
  {\bibfnamefont {F.}~\bibnamefont {Sanfilippo}}, \ and\ \bibinfo {author}
  {\bibfnamefont {A.}~\bibnamefont {Todaro}},\ }\href {\doibase
  10.1007/JHEP11(2018)170} {\bibfield  {journal} {\bibinfo  {journal} {JHEP}\
  }\textbf {\bibinfo {volume} {11}},\ \bibinfo {pages} {170} (\bibinfo {year}
  {2018})},\ \Eprint {http://arxiv.org/abs/1807.07954} {arXiv:1807.07954
  [hep-lat]} \BibitemShut {NoStop}%
\bibitem [{\citenamefont {Petreczky}\ \emph {et~al.}(2016)\citenamefont
  {Petreczky}, \citenamefont {Schadler},\ and\ \citenamefont
  {Sharma}}]{Petreczky:2016vrs}%
  \BibitemOpen
  \bibfield  {author} {\bibinfo {author} {\bibfnamefont {P.}~\bibnamefont
  {Petreczky}}, \bibinfo {author} {\bibfnamefont {H.-P.}\ \bibnamefont
  {Schadler}}, \ and\ \bibinfo {author} {\bibfnamefont {S.}~\bibnamefont
  {Sharma}},\ }\href {\doibase 10.1016/j.physletb.2016.09.063} {\bibfield
  {journal} {\bibinfo  {journal} {Phys. Lett.}\ }\textbf {\bibinfo {volume}
  {B762}},\ \bibinfo {pages} {498} (\bibinfo {year} {2016})},\ \Eprint
  {http://arxiv.org/abs/1606.03145} {arXiv:1606.03145 [hep-lat]} \BibitemShut
  {NoStop}%
\bibitem [{\citenamefont {Burger}\ \emph {et~al.}(2018)\citenamefont {Burger},
  \citenamefont {Ilgenfritz}, \citenamefont {Lombardo},\ and\ \citenamefont
  {Trunin}}]{Burger:2018fvb}%
  \BibitemOpen
  \bibfield  {author} {\bibinfo {author} {\bibfnamefont {F.}~\bibnamefont
  {Burger}}, \bibinfo {author} {\bibfnamefont {E.-M.}\ \bibnamefont
  {Ilgenfritz}}, \bibinfo {author} {\bibfnamefont {M.~P.}\ \bibnamefont
  {Lombardo}}, \ and\ \bibinfo {author} {\bibfnamefont {A.}~\bibnamefont
  {Trunin}},\ }\href {\doibase 10.1103/PhysRevD.98.094501} {\bibfield
  {journal} {\bibinfo  {journal} {Phys. Rev.}\ }\textbf {\bibinfo {volume}
  {D98}},\ \bibinfo {pages} {094501} (\bibinfo {year} {2018})},\ \Eprint
  {http://arxiv.org/abs/1805.06001} {arXiv:1805.06001 [hep-lat]} \BibitemShut
  {NoStop}%
\bibitem [{\citenamefont {Gorghetto}\ and\ \citenamefont
  {Villadoro}(2019)}]{Gorghetto:2018ocs}%
  \BibitemOpen
  \bibfield  {author} {\bibinfo {author} {\bibfnamefont {M.}~\bibnamefont
  {Gorghetto}}\ and\ \bibinfo {author} {\bibfnamefont {G.}~\bibnamefont
  {Villadoro}},\ }\href {\doibase 10.1007/JHEP03(2019)033} {\bibfield
  {journal} {\bibinfo  {journal} {JHEP}\ }\textbf {\bibinfo {volume} {03}},\
  \bibinfo {pages} {033} (\bibinfo {year} {2019})},\ \Eprint
  {http://arxiv.org/abs/1812.01008} {arXiv:1812.01008 [hep-ph]} \BibitemShut
  {NoStop}%
\bibitem [{\citenamefont {Dine}\ and\ \citenamefont
  {Kusenko}(2003)}]{Dine:2003ax}%
  \BibitemOpen
  \bibfield  {author} {\bibinfo {author} {\bibfnamefont {M.}~\bibnamefont
  {Dine}}\ and\ \bibinfo {author} {\bibfnamefont {A.}~\bibnamefont {Kusenko}},\
  }\href {\doibase 10.1103/RevModPhys.76.1} {\bibfield  {journal} {\bibinfo
  {journal} {Rev. Mod. Phys.}\ }\textbf {\bibinfo {volume} {76}},\ \bibinfo
  {pages} {1} (\bibinfo {year} {2003})},\ \Eprint
  {http://arxiv.org/abs/hep-ph/0303065} {arXiv:hep-ph/0303065 [hep-ph]}
  \BibitemShut {NoStop}%
\bibitem [{\citenamefont {Akita}\ \emph {et~al.}(2017)\citenamefont {Akita},
  \citenamefont {Kobayashi},\ and\ \citenamefont {Otsuka}}]{Akita:2017ecc}%
  \BibitemOpen
  \bibfield  {author} {\bibinfo {author} {\bibfnamefont {K.}~\bibnamefont
  {Akita}}, \bibinfo {author} {\bibfnamefont {T.}~\bibnamefont {Kobayashi}}, \
  and\ \bibinfo {author} {\bibfnamefont {H.}~\bibnamefont {Otsuka}},\ }\href
  {\doibase 10.1088/1475-7516/2017/04/042} {\bibfield  {journal} {\bibinfo
  {journal} {JCAP}\ }\textbf {\bibinfo {volume} {1704}},\ \bibinfo {pages}
  {042} (\bibinfo {year} {2017})},\ \Eprint {http://arxiv.org/abs/1702.01604}
  {arXiv:1702.01604 [hep-ph]} \BibitemShut {NoStop}%
\bibitem [{\citenamefont {Akita}\ and\ \citenamefont
  {Otsuka}(2019)}]{Akita:2018zma}%
  \BibitemOpen
  \bibfield  {author} {\bibinfo {author} {\bibfnamefont {K.}~\bibnamefont
  {Akita}}\ and\ \bibinfo {author} {\bibfnamefont {H.}~\bibnamefont {Otsuka}},\
  }\href {\doibase 10.1103/PhysRevD.99.055035} {\bibfield  {journal} {\bibinfo
  {journal} {Phys. Rev.}\ }\textbf {\bibinfo {volume} {D99}},\ \bibinfo {pages}
  {055035} (\bibinfo {year} {2019})},\ \Eprint
  {http://arxiv.org/abs/1809.04361} {arXiv:1809.04361 [hep-ph]} \BibitemShut
  {NoStop}%
\bibitem [{\citenamefont {Visinelli}\ and\ \citenamefont
  {Gondolo}(2009)}]{Visinelli:2009zm}%
  \BibitemOpen
  \bibfield  {author} {\bibinfo {author} {\bibfnamefont {L.}~\bibnamefont
  {Visinelli}}\ and\ \bibinfo {author} {\bibfnamefont {P.}~\bibnamefont
  {Gondolo}},\ }\href {\doibase 10.1103/PhysRevD.80.035024} {\bibfield
  {journal} {\bibinfo  {journal} {Phys. Rev.}\ }\textbf {\bibinfo {volume}
  {D80}},\ \bibinfo {pages} {035024} (\bibinfo {year} {2009})},\ \Eprint
  {http://arxiv.org/abs/0903.4377} {arXiv:0903.4377 [astro-ph.CO]} \BibitemShut
  {NoStop}%
\bibitem [{\citenamefont {Diez-Tejedor}\ and\ \citenamefont
  {Marsh}(2017)}]{Diez-Tejedor:2017ivd}%
  \BibitemOpen
  \bibfield  {author} {\bibinfo {author} {\bibfnamefont {A.}~\bibnamefont
  {Diez-Tejedor}}\ and\ \bibinfo {author} {\bibfnamefont {D.~J.~E.}\
  \bibnamefont {Marsh}},\ }\href@noop {} {\  (\bibinfo {year} {2017})},\
  \Eprint {http://arxiv.org/abs/1702.02116} {arXiv:1702.02116 [hep-ph]}
  \BibitemShut {NoStop}%
\bibitem [{\citenamefont {Chang}\ and\ \citenamefont
  {Cui}(2019)}]{Chang:2019mza}%
  \BibitemOpen
  \bibfield  {author} {\bibinfo {author} {\bibfnamefont {C.-F.}\ \bibnamefont
  {Chang}}\ and\ \bibinfo {author} {\bibfnamefont {Y.}~\bibnamefont {Cui}},\
  }\href@noop {} {\  (\bibinfo {year} {2019})},\ \Eprint
  {http://arxiv.org/abs/1910.04781} {arXiv:1910.04781 [hep-ph]} \BibitemShut
  {NoStop}%
\bibitem [{\citenamefont {Chang}\ and\ \citenamefont {Cui}(tion)}]{furture}%
  \BibitemOpen
  \bibfield  {author} {\bibinfo {author} {\bibfnamefont {C.-F.}\ \bibnamefont
  {Chang}}\ and\ \bibinfo {author} {\bibfnamefont {Y.}~\bibnamefont {Cui}},\
  }\href@noop {} {\  (\bibinfo {year} {In preparation})}\BibitemShut {NoStop}%
\bibitem [{\citenamefont {Salati}(2003)}]{Salati:2002md}%
  \BibitemOpen
  \bibfield  {author} {\bibinfo {author} {\bibfnamefont {P.}~\bibnamefont
  {Salati}},\ }\href {\doibase 10.1016/j.physletb.2003.07.073} {\bibfield
  {journal} {\bibinfo  {journal} {Phys. Lett.}\ }\textbf {\bibinfo {volume}
  {B571}},\ \bibinfo {pages} {121} (\bibinfo {year} {2003})},\ \Eprint
  {http://arxiv.org/abs/astro-ph/0207396} {arXiv:astro-ph/0207396 [astro-ph]}
  \BibitemShut {NoStop}%
\bibitem [{\citenamefont {Liddle}\ and\ \citenamefont
  {Scherrer}(1999)}]{Liddle:1998xm}%
  \BibitemOpen
  \bibfield  {author} {\bibinfo {author} {\bibfnamefont {A.~R.}\ \bibnamefont
  {Liddle}}\ and\ \bibinfo {author} {\bibfnamefont {R.~J.}\ \bibnamefont
  {Scherrer}},\ }\href {\doibase 10.1103/PhysRevD.59.023509} {\bibfield
  {journal} {\bibinfo  {journal} {Phys. Rev.}\ }\textbf {\bibinfo {volume}
  {D59}},\ \bibinfo {pages} {023509} (\bibinfo {year} {1999})},\ \Eprint
  {http://arxiv.org/abs/astro-ph/9809272} {arXiv:astro-ph/9809272 [astro-ph]}
  \BibitemShut {NoStop}%
\bibitem [{\citenamefont {Chung}\ \emph {et~al.}(2007)\citenamefont {Chung},
  \citenamefont {Everett},\ and\ \citenamefont {Matchev}}]{Chung:2007vz}%
  \BibitemOpen
  \bibfield  {author} {\bibinfo {author} {\bibfnamefont {D.~J.~H.}\
  \bibnamefont {Chung}}, \bibinfo {author} {\bibfnamefont {L.~L.}\ \bibnamefont
  {Everett}}, \ and\ \bibinfo {author} {\bibfnamefont {K.~T.}\ \bibnamefont
  {Matchev}},\ }\href {\doibase 10.1103/PhysRevD.76.103530} {\bibfield
  {journal} {\bibinfo  {journal} {Phys. Rev.}\ }\textbf {\bibinfo {volume}
  {D76}},\ \bibinfo {pages} {103530} (\bibinfo {year} {2007})},\ \Eprint
  {http://arxiv.org/abs/0704.3285} {arXiv:0704.3285 [hep-ph]} \BibitemShut
  {NoStop}%
\bibitem [{\citenamefont {Poulin}\ \emph {et~al.}(2018)\citenamefont {Poulin},
  \citenamefont {Smith}, \citenamefont {Grin}, \citenamefont {Karwal},\ and\
  \citenamefont {Kamionkowski}}]{Poulin:2018dzj}%
  \BibitemOpen
  \bibfield  {author} {\bibinfo {author} {\bibfnamefont {V.}~\bibnamefont
  {Poulin}}, \bibinfo {author} {\bibfnamefont {T.~L.}\ \bibnamefont {Smith}},
  \bibinfo {author} {\bibfnamefont {D.}~\bibnamefont {Grin}}, \bibinfo {author}
  {\bibfnamefont {T.}~\bibnamefont {Karwal}}, \ and\ \bibinfo {author}
  {\bibfnamefont {M.}~\bibnamefont {Kamionkowski}},\ }\href {\doibase
  10.1103/PhysRevD.98.083525} {\bibfield  {journal} {\bibinfo  {journal} {Phys.
  Rev.}\ }\textbf {\bibinfo {volume} {D98}},\ \bibinfo {pages} {083525}
  (\bibinfo {year} {2018})},\ \Eprint {http://arxiv.org/abs/1806.10608}
  {arXiv:1806.10608 [astro-ph.CO]} \BibitemShut {NoStop}%
\end{thebibliography}%

\end{document}